\documentclass[useAMS,usenatbib,onecolumn]{mn2e}
\usepackage{natbib}
\usepackage[dvipdfmx]{graphicx}
\usepackage{epstopdf}
\usepackage{times}
\usepackage{rotate}
\usepackage{color,url}
\newcommand{\bm}[1]{\mbox{\boldmath$#1$}}
\newcommand{\simgt}{\lower.5ex\hbox{$\; \buildrel > \over \sim \;$}}
\newcommand{\simlt}{\lower.5ex\hbox{$\; \buildrel < \over \sim \;$}}
\newcommand{\ave}[1]{\left\langle #1\right\rangle}
\newcommand{\rmd}{\ensuremath{\mathrm{d}}}

\begin{document}
\title[Joint 
analysis
of cluster counts 
and WL power spectrum]{Joint 
analysis
of cluster number counts and weak lensing power spectrum to correct for
the super-sample covariance}
\author[M. Takada and D.~N.~Spergel] {Masahiro Takada$^1$ and David
N. Spergel$^{1,2}$\\ 
$^1$Kavli Institute for the Physics and
Mathematics of the Universe (Kavli IPMU, WPI), The University of Tokyo,
Chiba 277-8582, Japan\\ 
$^2$Department of Astrophysical Sciences,
Princeton University, Peyton Hall, Princeton NJ 08544, USA }
\date{\today}


\maketitle

\begin{abstract}
 A coherent over- or under-density contrast across a finite survey
volume causes an upward- or downward-fluctuation in the observed number
of halos.  This fluctuation in halo number adds a significant co-variant
scatter in the observed amplitudes of weak lensing power spectrum at
nonlinear, small scales -- the so-called super-sample variance or the
halo sample variance.  
In this
paper, we show that by measuring both the number counts of clusters and
the power spectrum in the same survey region, we can mitigate this loss
of information and significantly enhance the scientific return from the
upcoming surveys.

First, using the halo model approach, we derive the cross-correlation
between the halo number counts and the weak lensing power spectrum,
taking into account the super-sample covariance effect, which well
matches the distributions measured from 1000 realizations for a
$\Lambda$-dominated cold dark matter model.
Then we show that adding the observed number counts of massive halos with
$M\simgt 10^{14}M_\odot/h$ can significantly improve the information
content of weak lensing power spectrum, almost recovering the Gaussian
information 
up to $l_{\rm max}\simeq 1000$,
if the average mass profiles of the massive halos 
are
known, which can be estimated from stacked lensing.  When combined with
the halo number counts for $M > 3 $ or $ 1 \times 10^{14}~M_\odot/h$,
the improvement is up to a factor of 1.4 or 2 at 
$l_{\rm max}\simeq 1000$--2000, equivalent to a factor 2 or 4 times
larger survey volume, compared to the power spectrum measurement alone.
\end{abstract}

\section{Introduction}

The world astronomy community is about to embark on 
wide-area
galaxy
surveys that
aim 
to use large-scale structure probes
to study the origin of cosmic acceleration.
 These range from
ground-based imaging and spectroscopic surveys such as the Subaru Hyper
Suprime-Cam (HSC) Survey
\footnote{\url{http://www.naoj.org/Projects/HSC/index.html}}\citep[see
also][]{Miyazakietal:12}, the Dark Energy Survey (DES)
\footnote{\url{http://www.darkenergysurvey.org}}, the Kilo-Degrees
Survey (KIDS) \footnote{\url{http://www.astro-wise.org/projects/KIDS/}},
the LSST \footnote{\url{http://www.lsst.org/lsst/}}, 
the Baryon
Oscillation Spectrograph Survey (BOSS)
\footnote{\url{http://cosmology.lbl.gov/BOSS/}}, the Extended BOSS survey
(eBOSS) \footnote{\url{http://www.sdss3.org/future/eboss.php}}, the
BigBOSS \footnote{\url{http://bigboss.lbl.gov}}, and the Subaru Prime
Focus Spectrograph (PFS) Survey
\footnote{\url{http://sumire.ipmu.jp/pfs/intro.html}}\cite[see
also][]{Takadaetal:12} 
to space-based optical and near-infrared missions
such as the Euclid project
\footnote{\url{http://sci.esa.int/science-e/www/area/index.cfm?fareaid=102}}
and the WFIRST project
\footnote{\url{http://wfirst.gsfc.nasa.gov}}\citep[see also][]{WFIRST}.
Each of these surveys approaches the nature of cosmic acceleration using
multiple large-scale structure probes: weak gravitational lensing,
baryon acoustic oscillations, clustering statistics of large-scale
structure tracers such as galaxies and clusters, the redshift-space
distortion effects, and the abundance of massive clusters
\citep[see][for a recent review]{Weinbergetal:12}.

Among the cosmological probes, 
weak lensing  measurements directly trace the distribution
of matter in the universe without assumptions about galaxy biases
and redshift space distortions
\citep[see][for a
review]{BartelmannSchneider:01}.  They are potentially the most powerful
cosmological probe in the coming decade 
\citep{Hu:99,Huterer:02,TakadaJain:04}.  Recent results such as  the Planck
lensing measurement \citep{PlanckLens:13} and the CFHT Lens Survey
\citep{Heymansetal:13,Kilbingeretal:13} are demonstrating the growing
power of these measurements.  

Most of the useful weak lensing signals 
are in the
nonlinear clustering regime, over the range of multipoles around
$l\simeq $ a few thousands \citep{JainSeljak:97,HutererTakada:05}.
Due
to mode-coupling nature of the nonlinear structure formation, the weak
lensing field at angular scales of interest  display large non-Gaussian
features. Thus, the two-point correlation function or the
Fourier-transformed counterpart, power spectrum,  no longer fully
describes the statistical properties of the weak lensing field.  Using ray-tracing
simulations and analytical methods such as the halo model approach,
previous works have shown that
the non-Gaussianity due to nonlinear structure formation
causes significant correlations between the power spectra at different multipoles
\citep{Jainetal:00,WhiteHu:00,CoorayHu:01,Sembolonietal:07,Satoetal:09,TakadaJain:09,Kiesslingetal:11,Kayoetal:13,KayoTakada:13}. In
particular, \citet{Satoetal:09} studied the power spectrum covariance
using 1000 ray-tracing simulation realizations, and showed that the non-Gaussian
error covariance degrades the information content by a factor of 2--3
for multipoles of a few thousands compared to the Gaussian information
of 
the initial density field.

What is the source of this non-Gaussian covariance that ``loses'' so
much of the hard-earned information in both the lensing power spectrum and galaxy redshift surveys?
\citet{Satoetal:09} \citep[see
also][]{TakadaJain:09,Kayoetal:13,TakadaHu:13,KayoTakada:13,Lietal:14} 
showed that
{\em super-sample variance} due to super-survey modes of length scales
comparable with or greater than a survey size is the leading source of
non-Gaussian covariance \citep[see also][for the pioneer
work]{Hamiltonetal:06}\footnote{Recently \citet{TakadaHu:13} developed
a unified theory of the power spectrum covariance including both the
weakly or deeply nonlinear versions of super-sample variance, which are
called beat coupling 
\citep{Hamiltonetal:06}  or  halo sample variance
\citep{Satoetal:09}, derived
based on the perturbation theory
or the halo model approach, respectively. For angular scales of interest in this
paper, the halo sample variance gives a dominant contribution to the
sample variance \citep{TakadaJain:09}. For this reason, 
we will often refer to the
halo sample variance as the super-sample variance in this paper.}.
The nonlinear version of the super-sample
variance can be physically interpreted as follows. If a survey region is
embedded in a coherent over- or under-density region, the abundance of
massive halos is up- or down-scattered from the ensemble-averaged
expectation as interpreted via halo bias theory
\citep{MoWhite:96,Moetal:97,ShethTormen:99} \citep[see
also][for the derivation of the super-sample variance of the halo number
counts]{HuKravtsov:03,HuCohn:06}. Then the modulation of halo abundance
in turn causes upward- and downward-fluctuations
in the amplitudes of weak lensing
power spectrum measured from the same survey region
\citep{TakadaBridle:07,Satoetal:09,Kayoetal:13}.  For angular scales
ranging from $l\sim 100$ to a few thousands, massive halos with $M\simgt
10^{14}M_\odot/h$ give a dominant contribution to the super-sample
variance of lensing power spectrum.

The information lost in
the power spectrum measurement can be recovered through measurements of
higher-order
correlation functions of the weak lensing field.  They 
add complementary information
that cannot be extracted by the power spectrum, even if measured from
the same survey region
\citep{TakadaJain:03,TakadaJain:03a,Sembolonietal:11,TakadaJain:04,Kayoetal:13,SatoNishimichi:13,KayoTakada:13}. 
There have also been a number of different approaches suggested for extracting
this complementary information: (1) performing a
 nonlinear transformation of the weak lensing
field and then studying  the power spectrum of the transformed field 
\citep{Neyrincketal:09,Seoetal:11,Zhangetal:11,Yuetal:12,Joachimietal:11,Seoetal:12}; or (2) 
using the statistics of rare peaks in the weak lensing mass map
\citep{Miyazakietal:02,Hamanaetal:04,Kratochviletal:10,Munshietal:12,Shirasakietal:12,Hamanaetal:12}.

Inspired by these previous works, the purpose of this paper is to study
a method of combining 
the abundance
of massive halos with the weak lensing power spectrum, 
in order to reduce the super-sample variance
contamination. Massive halos of $M\simgt 10^{14}M_\odot/h$ are
relatively easy to identify through a number of techniques such as identifying a
concentration of member galaxies in multi-color data
\citep{Rykoffetal:13} or identifying peaks 
in $X$-ray observations or in 
high-angular-resolution microwave surveys.  
By comparing the observed abundance of massive halos in the
survey region with the expectation for a fiducial cosmological model, we
can infer the effect of super-survey modes and therefore improve the
weak lensing power spectrum measurement. Based on this motivation, we
will first derive the covariance between the weak lensing
power spectrum and the number counts of massive halos for a given survey
region, using a method to model the likelihood function of halo number
counts \citep{HuCohn:06} and the halo model approach \citep[see
also][for the similar-idea study]{TakadaBridle:07}. Then, assuming that
the observed number counts of massive halos is available, we propose a
method of suppressing the 1-halo term contribution of the massive halos
to the weak lensing power spectrum measurement -- a Gaussianization
method. We will study how upcoming wide-area imaging surveys allow us to
implement the Gaussianization method in order to recover the information
content of the weak lensing power spectrum, compared to the maximum
information content of the initial Gaussian density field.

The structure of this paper is as follows. In Section~\ref{sec:idea}, we
motivate the method in this
paper.  In Section~\ref{sec:formulation}, we describe a formulation to
model the joint likelihood function of the halo number counts and the
matter power spectrum when both the two observables are drawn from the
same survey volume.  Then we discuss a ``Gaussianization'' method of
matter power spectrum estimation, which is feasible by combining with
the number counts of massive halos. In
Section~\ref{sec:wl}, we apply the formulas to the weak lensing power
spectrum measurement, assuming that massive halos in the surveyed
light-cone volume are identified. Assuming survey parameters for
upcoming wide-area galaxy surveys, we show how the Gaussianization
method of suppressing the 1-halo term contribution of massive halos can
recover the information content of the weak lensing power spectrum.
Section~\ref{sec:conclusion} is devoted to discussion and
conclusion. Unless explicitly denoted, we employ a $\Lambda$-dominated
cold dark matter ($\Lambda$CDM) model that is consistent with the WMAP
results.

\section{Basic idea}
\label{sec:idea}

\begin{figure}
\centering
\includegraphics[width=0.5\textwidth,angle=-90]{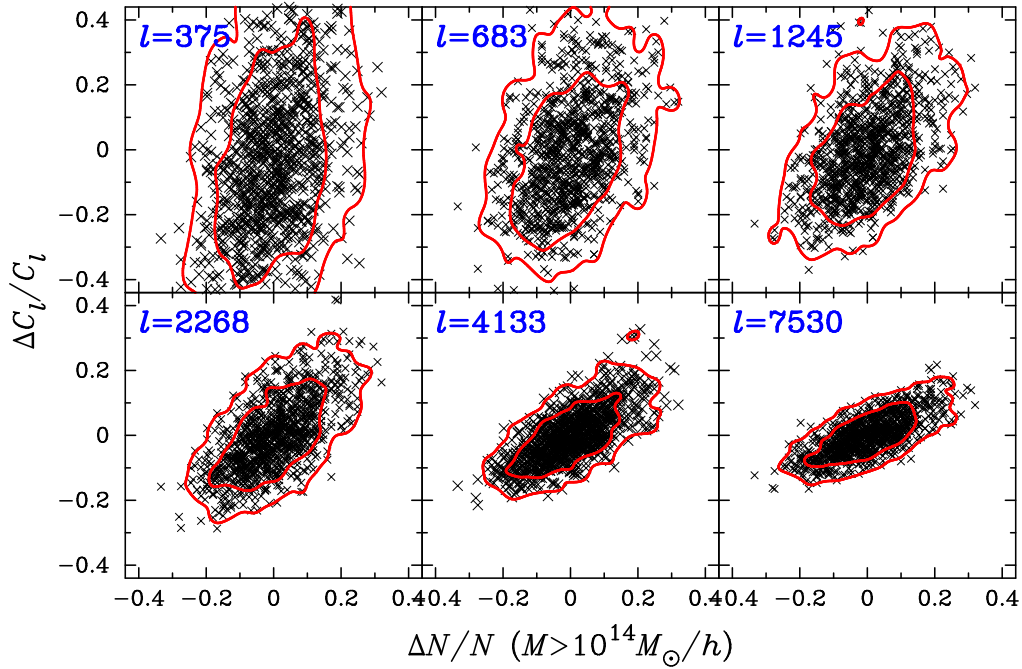}
\caption{The cross-correlation between the halo number counts and the
amplitudes of lensing power spectrum at different multipole bins,
measured from 1000 ray-tracing realizations for a $\Lambda$CDM model,
each of which has an area of 25 square degrees and has contributions of
super-survey modes (the $N$-body simulations used have the projected
angular scale greater than the ray-tracing area, $5$ degrees on a side).  
The halo
counts in the $x$-axis is for halos with masses greater than
$10^{14}M_\odot/h$. The different cross symbols are for different
realizations, and the red solid contours show  68 or 95 percentile
regions that are computed by smoothing the distribution with a
two-dimensional Gaussian kernel that has widths of 1/50th the plotted
ranges in the $x$- and $y$-axes. Note that the plotted ranges of $x$-
and $y$-axes are the same in all the panels.
For multipole bins
around $l\simeq 1000-3000$, 
the power
spectrum amplitudes are highly correlated with the number counts of
halos with $M\ge 10^{14}M_\odot$.
\label{fig:dn-dcl}
} 
\end{figure}
\begin{figure}
\centering
\includegraphics[width=0.44\textwidth,angle=-90]{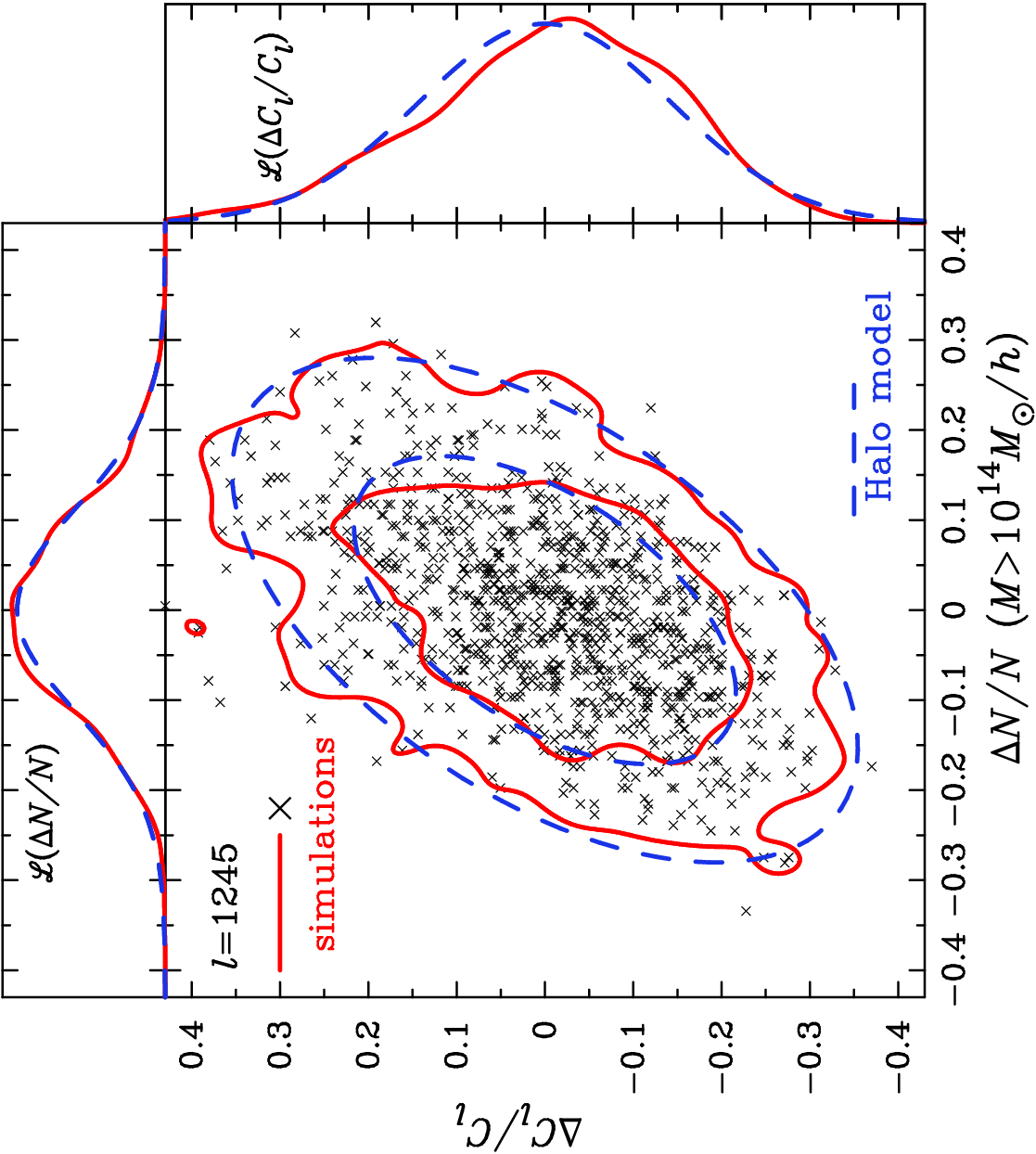}
\includegraphics[width=0.44\textwidth,angle=-90]{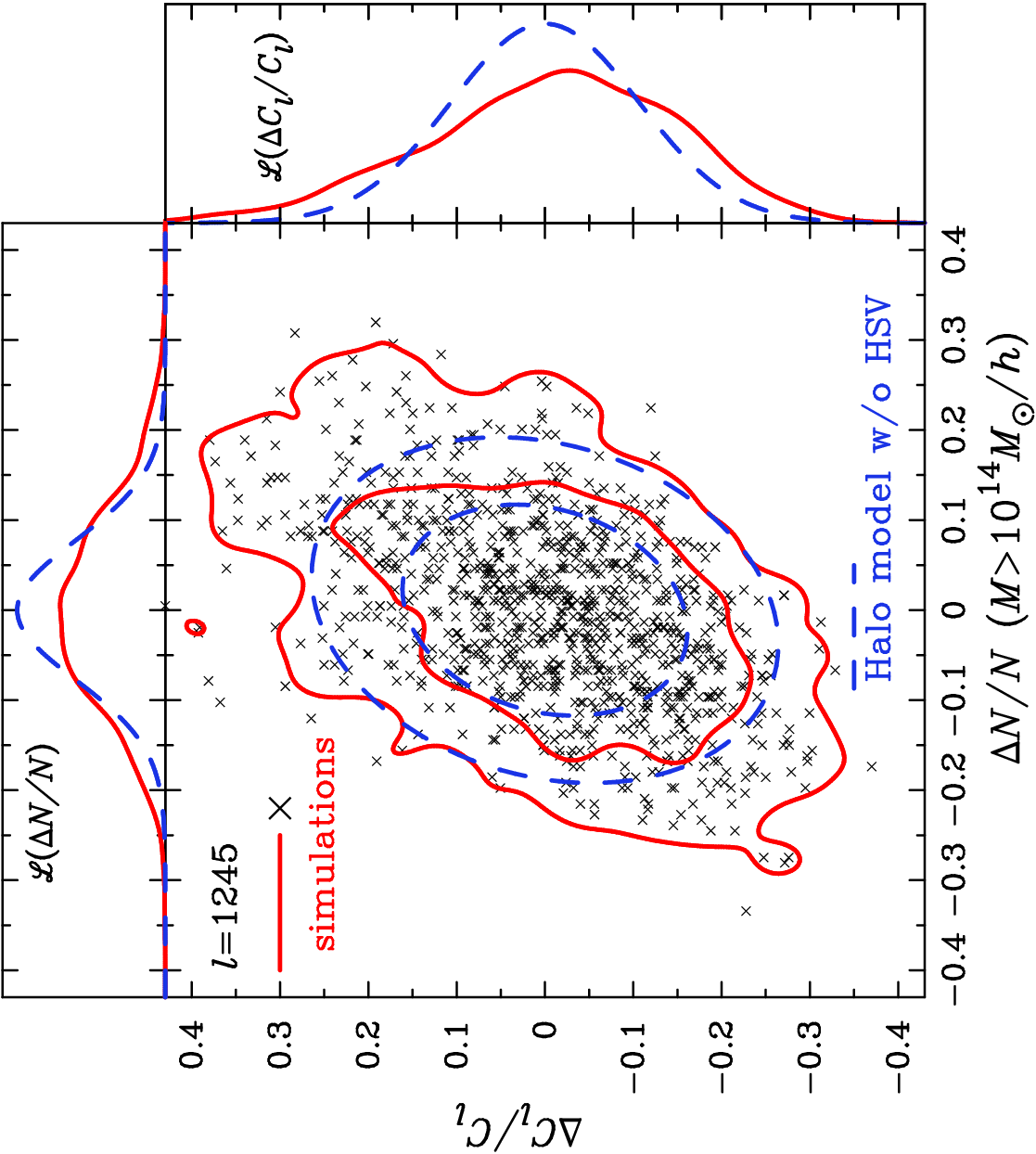}
\caption{This plot compares the
halo model predictions (Eqs.~\ref{eq:joint_n-wl}--\ref{eq:cov_n-wl})
to 
the results of the numerical simulations shown l in the previous figure for the 
multipole bin centered at $l=1245$). {\em Left panel}:The
 model includes the halo sampling variance (HSV) contribution, which
 arises from the super-survey modes of scales comparable with or outside
 the survey region.  With the HSV term, the halo model predictions agrees with the simulation results. The 
 upper- or left-side panels
 show the projected one-dimensional likelihood functions for each
 observable, and again shows that the halo model predictions well
 reproduce the simulation results.  {\em Right}: The model predictions
 without the HSV cannot reproduce the simulation results. The plot shows
 a only weak correlation between the two observables. 
These results suggest that the number counts of the massive halos in a
given survey region can be used to correct for the HSV contribution to
the power spectrum measurement.  \label{fig:dn-dcl-model} }
\end{figure}

There are several sources  of statistical fluctuations in measurement of the 
weak lensing power spectrum and the number counts of halos.
The finite size of a survey implies that there will be  Poisson noise in the halo
number counts and cosmic variance due to the finite number of Fourier modes available for the
power spectrum measurement. 
 {\em
Super-sample variance} \citep{Hamiltonetal:06,TakadaHu:13}
or the halo sample variance (HSV)
\citep{Satoetal:09,Kayoetal:13}
is an important additional source of statistical fluctuations.  
 \citet{Satoetal:09} \citep[see also][]{Kayoetal:13} showed that the HSV gives a significant contribution
to the power spectrum covariance  at  $l\simgt 1000$.

Numerical simulations clearly show the coherent fluctuation in the power
spectrum due to halo sample variance.  Fig.~\ref{fig:dn-dcl} shows how
the number of massive halos in a light-cone volume is correlated with
the amplitude of lensing power spectrum measured from the same volume.
For this figure (and for other analyses in this paper), we used 1000
simulation realizations, generated in \cite{Satoetal:09} \citep[see
also][]{Satoetal:11}, each of which has an area of 25 square degrees and
contains both the distribution of massive halos and the lensing field
for source galaxies at redshift $z_s=1$.  The ray-tracing simulations
are done in a light-cone volume, and have contributions from the
super-survey modes, because the $N$-body simulations used for modeling
the nonlinear large-scale structure contain the modes of projected
length scales greater than the light-cone size (5~degrees on a side) at
each lens redshift bin \citep[see Fig.~1 in][]{Satoetal:09}. Hence, with
the ray-tracing simulations, we can study the effect of super-survey
variance on the halo number counts and the power spectrum estimation.
For the number counts of halos, we included massive halos with masses
$M\ge 10^{14}M_\odot$.  The cross symbols in each panel denote the
different realizations, and the solid contours show 68 or 95 percentile
regions of the distribution. Shown here is the fractional variations of
the two observables, where the quantities in the denominator,
$\bar{C}_l$ or $\bar{N}$, are their mean values among the 1000
realizations. The two observables are highly correlated with each other
at high multipole bins, $\ell \simgt 1000$. For massive halos with $M\ge
10^{14}M_\odot/h$, the spectrum amplitude of multipole bin centered at
$l=1245$ shows a strongest correlation with the number counts,
displaying an almost linear relation of $\Delta N/\bar{N}\propto \Delta
C_l/\bar{C}_l$.

In Fig.~\ref{fig:dn-dcl-model}, we demonstrate that the analytical
model developed in Sections~\ref{sec:formulation} and \ref{sec:wl} 
reproduces  the
simulation result in Fig.~\ref{fig:dn-dcl}.  To compute the model
predictions, we assume a multivariate Gaussian distribution of the two
observables, assuming that their widths and the cross-correlation
strength are given by the covariances and the cross-covariance computed
based on the halo model, as we will develop in Section~\ref{sec:wl}. 
The right panel explicitly shows that, if the
HSV effect is ignored, the model predicts a only weak correlation
between the two observables, which does not match the simulation result.

The results in Figs.~\ref{fig:dn-dcl} and \ref{fig:dn-dcl-model} imply
that, by using the observed number counts of massive halos in each
survey volume, one can calibrate or correct for the super-sample
variance effect
in the power spectrum estimation. 
This is the question that we address in the
following sections.

\section{Formulation: Covariance of halo number counts and 
matter power spectrum}
\label{sec:formulation}

\subsection{Likelihood function of halo number counts}

In this section, we briefly review the likelihood function of cluster
number counts taking into account the super-sample variance, based on
the method developed in \citet{HuKravtsov:03}, \citet{HuCohn:06} and
\cite{TakadaBridle:07}.

Consider a finite-volume survey of comoving volume $V_s$ that has
 an over- or under-density given by
\begin{equation}
\bar\delta_m(V_s)\equiv \int\!\rm.^3\bm{x}~ 
\delta_m(\bm{x})W(\bm{x}; V_s),
\end{equation}
where $W(\bm{x}; V_s)$ is the survey window function;
$W(\bm{x})=1$ if $\bm{x}$ is inside a survey region, otherwise
$W(\bm{x})=1$, and is
defined so as to
satisfy the normalization condition $\int\!\!\rmd^3\bm{x}W(\bm{x})=1$.
  We can use 
 halo bias theory \citep{MoWhite:96,ShethTormen:99} to estimate how this super-survey
mode modulates the predicted
number counts of halos in mass range $[M,M+dM]$  from its
ensemble average expectation:
\begin{equation}
\bar{N}(M)=V_s\frac{\rmd n}{\rmd M}\rmd M \rightarrow
 N(M)=V_s\frac{\rmd n}{\rmd M}\rmd M\left[1+b(M)\bar{\delta}_m(V_s)\right],
\label{eq:n_mod}
\end{equation}
where $\rmd n/\rmd M$ is the halo mass function, 
$V_s(\rmd n/\rmd M)\rmd M$ is 
an ensemble average expectation
of the number counts,
and $b(M)$ is the halo bias. Note that we use the
same model ingredients in \cite{OguriTakada:11} to compute these
quantities for a given cosmological model.  

For a sufficiently large volume in which we are most interested, the
density fluctuation $\bar\delta_m$ is considered to be well in the
linear regime, and the probability distribution of $\bar\delta_m$ is
approximated by  a Gaussian distribution:
\begin{equation}
P(\bar\delta_m)=\frac{1}{\sqrt{2\pi}\sigma_m(V_s)}\exp\left[
-\frac{\bar\delta_m^2}{2\sigma_m^2(V_s)}\right].
\label{eq:gauss_dm}
\end{equation}
The variance $\sigma_m(V_s)$ is the rms mass density fluctuations
of the survey volume $V_s$, defined in terms of the linear mass
power spectrum as
\begin{equation}
\sigma^2_m(V_s)\equiv \int\!\!\frac{\rmd^3\bm{k}}{(2\pi)^3}~P^L_m(k)\left|
\tilde{W}(\bm{k}; V_s) 
\right|^2,
\end{equation}
where $P^L_m(k)$ is the linear matter power spectrum, and
$\tilde{W}(\bm{k})$ is the Fourier transform of the survey window
function (the window function is generally
anisotropic in Fourier space, depending on the geometry of survey
region).
$\bar\delta_m$ and $\sigma_m$ have
contributions from Fourier modes of scales comparable with or outside
the survey volume, so are therefore {\em not} a direct observable.

We next construct  an estimator of the number counts of halos in
different mass bins:
$\hat N_1, \hat N_2, ...$, and $\hat N_k$ 
in mass bins of
$M_1, M_2, ...$ and $M_k$, respectively. Assuming a joint Poisson distribution,
the joint probability
distribution 
is given as 
\begin{equation}
{\cal L}(\hat N_1, \hat N_2,..., \hat N_k; \bar\delta_m) =
\prod_{i=1}^{k}
\frac{N_i^{\hat N_i}}{\hat N_i!} \exp(-N_i),
\label{eq:p_nd}
\end{equation}
where 
\begin{equation}
N_i\equiv \bar{N}_i[1+b_i \bar\delta_m(V_s)],
\end{equation}
 $\bar{N}_i \equiv V_s(\rmd n/\rmd M_i)dM_i$ and $b_i\equiv 
b(M_i)$.  In the following,
quantities with hat symbol ``$\hat{~ }$'' denote
estimators or observables that can be estimated from a survey, 
 and the quantities with bar symbol ``$\bar{~ }$'', except for
 $\bar\delta_m(V_s)$,  
denote the
ensemble-average expectation values. 

Since $\bar{\delta}_m(V_s)\ll 1$ for a case we are interested in, 
expanding the likelihood function (Eq.~\ref{eq:p_nd}) to second order in
$\bar\delta_m$ yields
\begin{eqnarray}
{\cal L}(\hat N_1, \hat N_2,...,\hat N_k; \bar{\delta}_m) &\simeq & 
\left[
\prod_{i=1}^k \frac{\bar{N}_i^{\hat N_i}}{\hat N_i!}\exp(-\bar{N}_i) 
\right] 
\left(1 + \sum_j b_j \hat N_j \bar\delta_m + \sum_{j,j'}b_j \hat N_j b_{j'}
 \hat N_{j'}\bar{\delta}_m^2
+ \sum_j \frac{b_j\hat N_j (b_j\hat N_j-1)}{2} \bar\delta_m^2 \right) \nonumber \\ 
&& \times \left(1 -\sum_j \bar{N}_j b_j \bar\delta_m + 
\sum_{j,j'}\bar{N}_j b_j \bar{N}_{j'}b_{j'} \bar{\delta}_m^2+
\frac{1}{2}\sum_j\bar{N}_j^2 b_j^2 \bar\delta_m^2\right)\nonumber \\
&&\hspace{-2em}\simeq
\left[
\prod_{i=1}^k \frac{\bar{N}_i^{\hat N_i}}{\hat N_i!}\exp(-\bar{N}_i) 
\right]
\left[
1+\sum_{j}\left(b_j\hat N_j-\bar{N}_jb_j\right)\bar\delta_m \right.
\nonumber\\
&&\hspace{-0em}+
\left.
\left\{
\sum_{j,j'}\left(b_j\hat N_jb_{j'}\hat N_{j'}-b_j\hat N_jb_{j'}\bar{N}_{j'}+b_{j}\bar{N}_j
b_{j'}\bar{N}_{j'}
\right)
+\frac{1}{2}\sum_j\bar{N}_j^2 b_j^2 
+\sum_{j}\frac{b_j\hat N_j(b_j\hat N_j-1)}{2}\right\}\bar\delta_m^2
\right].
\end{eqnarray}
By integrating over the density contrast $\bar\delta_m$ with its
probability distribution (Eq.~\ref{eq:gauss_dm}), we can derive the
joint probability distribution for the number counts of halos that include
marginalizing  over the amplitude of  the super-survey mode
$\bar\delta_m$:
\begin{eqnarray}
{\cal L}(\hat N_1,\hat N_2,\dots, \hat N_k)&=&
\left[
\prod_{i=1}^k \frac{\bar{N}_i^{\hat N_i}}{\hat N_i!}\exp(-\bar{N}_i) 
\right]
\left[
1+\frac{1}{2}\left\{
\left(\sum_jb_j(\hat N_j-\bar{N}_j)\right)^2-\sum_jb_j^2\hat N_j
\right\}
\sigma_m^2
\right].
\label{eq:p_app}
\end{eqnarray}
This is a slight generalization of Eq.~(16) in \cite{HuCohn:06}. Since
the quantities $\bar{N}_i$, $b_i$ and $\sigma_m^2$ can be computed 
once 
a cosmological model, the survey window function and the halo mass
bins are specified, 
we
can evaluate the joint probability distribution for the observed 
number counts $\left\{\hat{N}_i\right\}$ for the assumed cosmological model.
In practice,
we need to also include observational effects such as detector noise
and halo mass proxy uncertainty, but we do not consider the effects in
this paper for simplicity.

By using the probability distribution function (Eq.~\ref{eq:p_app}), we
can find the following summation rules for the halo number counts of a single
mass bin:
\begin{eqnarray}
&&\sum_{\hat N=0}^{\infty} {\cal L}(\hat N)=1, \nonumber\\
&&\ave{\hat N}\equiv \sum_{\hat N=0}^{\infty} \hat N{\cal L}(\hat N)=\bar{N}, \nonumber\\
&&\ave{\hat N^2}\equiv \sum_{\hat N=0}^{\infty} \hat N^2{\cal L}(\hat N)=\bar{N}+\bar{N}^2+b^2\bar{N}^2\sigma_m^2,
\label{eq:p_app_form1}
\end{eqnarray}
Note again $\bar{N}=V_s(\rmd n/\rmd M)\Delta M$, the ensemble-average expectation
value of the
number counts corresponding to the counts for an infinite-volume survey.  
Hence, the variance of the halo number counts
 is found to be
\begin{equation}
\sigma^2(\hat N)\equiv \ave{\hat N^2}-\ave{\hat{N}}^2=\bar{N}+
							       b^2\bar{N}^2\sigma_m^2.
\end{equation}
The first term is a Poisson noise contribution arising due to a finite
number of sampled halos. 
The second term is the halo sample variance (HSV) contribution arising
due to 
super-survey modes.
\citet{Crocceetal:10} showed that,
using cosmological simulations of a sufficiently
large volume, the above formula can accurately describe 
sample variances of
the halo number counts measured from subdivided volumes of $N$-body
simulation,
where the sub-volumes were considered in order to study the effect
of super-survey modes on the sample variance.

Next let us consider the joint probability distributions for the halo
number counts, $\hat{N}_i$ and $\hat{N}_j$, in two mass bins $M_i$ and
$M_j$, respectively ($i\ne j$). From Eq.~(\ref{eq:p_app}), we can find that the joint
distribution can be rewritten as 
\begin{eqnarray}
{\cal L}(\hat N_i,\hat N_j)={\cal L}(\hat N_i){\cal L}(\hat N_j)+
\frac{e^{-\bar{N}_i}\bar{N}_i^{\hat N_i}}{\hat N_i !}
\frac{e^{-\bar{N}_j}\bar{N}_j^{\hat N_j}}{\hat N_j !}
b_ib_j(\hat N_i-\bar{N}_i)(\hat N_j-\bar{N}_j)\sigma_m^2.
\label{eq:n_cov1}
\end{eqnarray}
Then we can find the following summation rules: 
\begin{eqnarray}
&&\sum_{\hat N_i=0}^{\infty}
\sum_{\hat N_j=0}^{\infty}{\cal L}(\hat N_i,\hat N_j)=1, \nonumber\\
&&\ave{\hat{N}_i\hat{N}_j}\equiv 
\sum_{\hat N_i=0}^{\infty}
\sum_{\hat N_j=0}^{\infty}{\cal L}(\hat N_i,\hat N_j)\hat N_i\hat N_j=\bar{N}_i\bar{N}_j
\left(1+b_ib_j\sigma_m^2\right), \hspace{2em} (i\ne j).
\label{eq:n_cov2}
\end{eqnarray}
Similarly, the variance reads
\begin{equation}
\sigma^2(\hat N_i\hat{N}_j)\equiv \ave{\hat N_i \hat
 N_j}-\ave{\hat N_i}\ave{\hat N_j} =b_ib_j \bar{N}_i\bar{N}_j 
\sigma_m^2. 
\end{equation}
Thus the number fluctuations in halos of two mass bins are positively
correlated with each other. The similar formula hold for more than three
bins. 

Combining Eqs.~(\ref{eq:n_cov1}) and (\ref{eq:n_cov2}), we can re-write
the ensemble average of the halo number counts as
\begin{equation}
\ave{\hat N_i\hat N_j}=\bar N_i\delta^K_{ij}+\bar N_i \bar
 N_j\left(1+b_ib_j
\sigma_m^2\right),
\label{eq:n_cov3}
\end{equation}
where $\delta^K_{ij}$ is the Kronecker delta function: $\delta_{ij}^K=1$
if $i=j$, otherwise $\delta^K_{ij}=0$.

Similarly, we can find
\begin{equation}
 \ave{\hat N_i \hat N_j \hat N_k}=\bar{N}_i\bar{N}_j\bar{N}_k
\left[
1+\left(
b_ib_j+b_jb_k + b_k b_i
\right)\sigma_m^2
\right],
\end{equation}
for $i\ne j, j\ne k$ and $k\ne i$.

\subsection{Matter power spectrum and the covariance matrix}

In this section, using the halo model formulation
\citep{Seljak:00,PeacockSmith:00,MaFry:00,Scoccimarroetal:01,TakadaJain:03a}
\citep[see also][for a review]{CooraySheth:02} as well as the joint likelihood of
halo number counts (Eq.~\ref{eq:p_app}), we derive the covariance matrix
for the three-dimensional matter power spectrum including the
super-sample variance contribution.

In the halo model approach, the matter
power spectrum is given by a
sum of the 1- and 2-halo terms that arise
from correlations of matter  within  the same one halo and between 
different halos, respectively:
\begin{eqnarray}
P^{1h}(k)&=&\int\!\rmd M\frac{\rmd n}{\rmd M} 
\left(
\frac{M}{\bar{\rho}_{m}}\right)^2
\left|\tilde{u}_M(k)\right|^2, \nonumber \\
P^{2h}(k)&=& 
\int\!\rmd M\frac{\rmd n}{\rmd M}\frac{M}{\bar{\rho}_m}
\int\!\rmd M'\frac{\rmd n}{\rmd M'}\frac{M'}{\bar{\rho}_m} P_{hh}(k; M, M'),
\label{eq:ps_halomodel}
\end{eqnarray}
where $\tilde{u}_{M}(k)$ is the Fourier transform of the average mass
profile of halos with mass $M$, and $P_{hh}(k;M, M')$ is the power
spectrum between two halos of masses $M$ and $M'$. Throughout this paper 
we assume an Navarro-Frenk-White
(NFW) halo mass profile \citep{Navarroetal:97}.  The factor $(M/\bar\rho_m)$ in
the above equation accounts for the fact that more massive halos contain
more dark matter particles.
We assume that the ensemble average of 
the halo power spectrum is given by 
\begin{equation}
P_{hh}(k; M, M')\simeq 
b(M)b(M')\tilde{u}_{M}(k)\tilde{u}_{M'}(k)P^{L}_m(k). 
\label{eq:ens_2h}
\end{equation} 
If we were working with real data, rather than assuming an NFW profile, we could measure the
halo profile by stacking clusters identified by other techniques.

If recalling that the halo number counts in a given mass range is given
as $N=V_s\left(\rmd n/\rmd M\right)\Delta M$ in an ensemble average
sense, Eq.~(\ref{eq:ps_halomodel}) leads us to define estimators of the
1- and 2-halo power spectra in terms of the observed halo number counts
as
\begin{eqnarray}
\hat{P}^{1h}(k)&=&\frac{1}{V_s}\sum_i \hat N_i
 \hat{p}_i^{1h}(k),\nonumber \\
\hat{P}^{2h}(k)&=&\frac{1}{V_s^2}\sum_{i,j} \hat N_i\hat N_j
\hat{p}_{ij}^{2h}(k),
\label{eq:ps1h-2h_est}
\end{eqnarray}
where $\hat N_i\equiv \hat N(M_i)$ and we have approximated the
integration in Eq.~(\ref{eq:ps_halomodel}) by a discrete summation over
different halo mass bins. $\hat{p}^{1h}_i(k)$ and $\hat{p}^{2h}_{ij}(k)$
are estimators that are given in terms of the mass density field. More
specifically, $\hat{p}^{1h}_i(k)$ arises from the matter distribution
inside halos of the $i$-th mass bin, $M_i$, and the ensemble average
gives the average mass profile of the halos. $\hat{p}^{2h}_{j}(k)$ is
from the mass field that governs clustering of different halos in mass
bins $M_i$ and $M_j$,
and the
ensemble average 
gives
the linear mass power spectrum,
weighted with the halo biases $b_i$ and $b_j$
(see
Eq.~\ref{eq:ens_2h}). 
We assume that, since the super-survey mode,
$\bar{\delta}_m$, contributes only to the monopole of the Fourier modes
in a finite survey region (that is, $\bar{\delta}_m$ is a constant,
background mode across the survey volume),
$\bar{\delta}_m$ is not correlated with 
$\hat{p}_i^{1h}(k)$ and $\hat{p}_{ij }^{2h}(k)$; $\ave{\bar\delta_m
\hat{p}^{1h}_i}= \ave{\bar\delta_m \hat{p}^{2h}_{ij}}=0$. 
In other words,
we
assume that the
super-survey mode affects the matter power spectrum $\hat{P}(k)$ only
through its effect on the halo number counts $\hat{N}_i$. 

As derived in 
detail in Appendix~\ref{app:pscov}, by using the joint
probability distribution function for the halo number counts,
${\cal L}(\hat N_1, \hat{N}_2, \dots)$ (Eq.~\ref{eq:p_app}), 
we can derive the power spectrum
covariance as
\begin{equation}
{\rm Cov}[\hat{P}(k),\hat{P}(k')]=\frac{2}{N_{\rm mode}(k)}P(k)^2\delta^K_{kk'}
+\frac{1}{V_s}\bar{T}(k,k')+\left[
\int\!\rmd M\frac{\rmd n}{\rmd M}b(M)p^{1h}_M(k)
\right]
\left[
\int\!\rmd M'\frac{\rmd n}{\rmd M'}b(M')p^{1h}_{M'}(k')
\right]\sigma_m^2(V_s).
\label{eq:pscov}
\end{equation}
Here $N_{\rm
mode}(k)$ is the number of independent Fourier modes centered at $k$,
where we mean by ``independent'' that the Fourier modes are
discriminated by the fundamental mode of a given survey, $k_f\simeq
2\pi/V_s^{1/3}$. For the case of $k\gg k_f$, 
\begin{equation}
N_{\rm mode}(k)\simeq \frac{4\pi k^2\Delta k}{k_f^3}=\frac{k^2\Delta k V_s}{2\pi^2},
\label{eq:Nmode}
\end{equation}
where $\Delta k$ is the bin width. 
Eq.~(\ref{eq:pscov}) reproduces Eq.~(11) in \citet{Kayoetal:13}. 
$\bar{T}(k,k')$ is the
angle-averaged trispectrum \citep[see around Eq.~14 in][]{Satoetal:09}.
The first and second terms on the r.h.s. of Eq.~(\ref{eq:pscov}) are
standard terms of the power spectrum covariance that have been
considered in the literature \citep{Scoccimarroetal:99}.  The terms both
scale with survey volume as $\propto 1/V_s$; a larger survey volume
reduces the amplitudes.  The third term of Eq.~(\ref{eq:pscov}) is the
super-sample variance or the HSV contribution
\citep{Satoetal:09,Kayoetal:13,TakadaHu:13}. Very similarly to the
effect on the halo number counts, a coherent over- or under-density
mode
in a given survey region
causes an upward or
downward scatter in the power spectrum amplitudes, respectively. 
At large $k$ limit,
where the 1-halo term is dominant, the HSV term 
behaves
like
${\rm Cov}[P(k),P(k')]_{\rm HSV}\propto
P^{1h}(k)P^{1h}(k')\sigma_m^2$ or the correlation coefficient
matrix
$r(k,k')\equiv 
{\rm Cov}[P(k),P(k')]_{\rm HSV}/[P(k)P(k')]\propto \sigma_m^2
$. 
That is, the HSV adds powers to the diagonal and off-diagonal
components of the covariance matrix in the exactly same way.
The
dependence of the HSV term on survey volume differs from 
other terms as it scales with survey volume via $\sigma_m$, which depends
on the linear mass power spectrum $P^L_m(k)$ convolved with 
the survey window
function that has a width of $1/L~ (V_s\sim L^3)$ in Fourier space
\citep[see Section~3.1 in][for the details]{Kayoetal:13}.  

\subsection{Cross-correlation between the halo number counts and the
  matter power spectrum}

Eq.~(\ref{eq:ps1h-2h_est}) implies that the power spectrum estimators
are correlated with the halo number counts \citep{TakadaBridle:07}, if
the two observables are drawn from the same survey region. Similarly, as
shown in 
Appendix~\ref{app:cross}, 
we derive the
cross-covariance between 
 the halo number counts of the
$i$-th mass bin, $M_i$ and the power spectrum amplitude
at the $k$-bin:
\begin{eqnarray}
{\rm Cov}[\hat{N}_i,\hat{P}(k)]&=& \frac{1}{V_s}\bar{N}_{i}p_{i}^{1h}(k)+
b_{i}\bar{N}_{i} \sigma_m^2\int\!\!\rmd M\frac{\rmd n}{\rmd M} b(M)p_M^{1h}(k)
\nonumber\\
&&
+\bar{N}_i\sigma_m^2
\int\!\rmd M\rmd M'\frac{\rmd n}{\rmd M}\frac{\rmd n}{\rmd M'}\left[
2b(M_i)b(M)+b(M)b(M')
\right]p^{2h}_{MM'}(k).
\label{eq:crosscov}
\end{eqnarray}
The second and third terms with $\sigma_m^2$ explicitly show that the
halo number counts is correlated with the power spectrum amplitudes
through the super-survey modes. Note that the above formula has a
similar form to that in \cite{TakadaBridle:07}.
The first term is negligible
compared to other terms if a mass bin of halos is sufficiently narrow. 

\subsection{A Gaussianized estimator of matter power spectrum:
   suppressing the 1-halo term contribution of massive halos }
\label{sec:dp}

As shown in \citet{Satoetal:09} and \cite{Kayoetal:13}, the non-Gaussian
errors significantly degrade the cumulative signal-to-noise ratio or the
information content of power spectrum measurement compared to the
Gaussian expectation which is originally contained in 
the initial density field of structure
formation. The degradation is significant in the
nonlinear regime, 
and is mainly from the HSV
contribution. 
In particular, for the nonlinear scales around
$k\simeq$ a few $h/{\rm Mpc}$ (corresponding to
 angular scales of $l\simeq 10^3$ for the weak lensing power spectrum), the
HSV effect arises mainly from massive halos with masses $M\simgt
10^{14}M_\odot/h$. Such massive halos are relatively easy to identify in
the survey region, e.g., from a concentration of member galaxies. These
suggest that, by combining the observed number counts of massive halos
with a measurement of power spectrum, we may be able to correct for the
HSV effect on the power spectrum -- a Gaussianization method of the
power spectrum measurement. 
In this section, we study this method. 
Note that this method is {\em not} feasible
if the two observables are drawn from {\em different} survey regions. 

First, let us consider an ideal case: suppose that we have a measurement
of the halo number counts $\left\{\hat{N}_2,\hat N_2, \dots,
\hat{N}_k\right\}$ in mass bins of $M_1, M_2, \dots, M_k$ for a given
survey of comoving volume $V_s$. Also suppose that we have an estimator of the
mass density field, $\delta_m(\bm{x})$, in order to estimate the matter
power spectrum for the same survey region.
For these assumptions, 
we can define an estimator of the matter
power spectrum with suppressing the 1-halo term contribution: 
\begin{equation}
\widehat{\Delta P(k)}=\hat{P}(k)-\frac{1}{V_s}\sum_i \hat N_i p^{1h}_i(k)
=\hat{P}^{2h}(k)+\frac{1}{V_s}\sum_i \hat N_i
\left[\hat{p}^{1h}_i(k)-p^{1h}_i(k )\right],
\label{eq:est_dpk}
\end{equation}
where $p^{1h}_i(k)$ is a theory template for the 1-halo term power
spectrum of halos in the $i$-th mass bin $M_i$, for an assumed
cosmological model (e.g.,  an NFW profile for the assumed cosmology).

The ensemble average of
the power spectrum estimator (Eq.~\ref{eq:est_dpk}) reads
\begin{eqnarray}
\ave{\widehat{\Delta P}(k)}&=&\ave{\hat{P}^{2h}(k)}+\frac{1}{V_s}\sum_i \ave{
\hat N_i\left[\hat{p}^{1h}_i(k)-p^{1h}_i(k)\right]
}\nonumber\\
&=&
 P^{2h}(k)+\frac{1}{V_s}\sum_i\ave{\hat N_i}\left[\ave{\hat{p}^{1h}(k)}-p^{1h}(k)\right]\nonumber\\
&=& P^{2h}(k), 
\end{eqnarray}
where we have assumed that  the 1-halo term power template
spectrum matches the underlying true spectrum after the ensemble
average.  Thus the ensemble average of the estimator (\ref{eq:est_dpk})
leaves only the 2-halo term. 
The covariance matrix for the estimator (\ref{eq:est_dpk})
is found from Eq.~(\ref{eq:pscov}) to be
\begin{eqnarray}
{\rm Cov}\left[\widehat{\Delta P}(k),\widehat{\Delta P}(k')
\right]\simeq \frac{2}{N_{\rm mode}(k)}\delta_{kk'}^K P^{2h}(k)^2.
\end{eqnarray}
The power spectrum estimator (Eq.~\ref{eq:est_dpk}) suppressing the
1-halo term contribution obeys a Gaussian error covariance. In other
words, it reduces the non-Gaussian errors including the HSV effect, and
can recover the Gaussian information content.

In reality,  we can only identify halos with
masses greater than a certain mass threshold $M_{\rm th}$.  Given this limitation,
 the power spectrum estimator with suppressing the 1-halo term
contribution needs to be modified from Eq.~(\ref{eq:est_dpk}) as
\begin{equation}
\widehat{\Delta P}(k)\equiv \hat{P}(k)-\frac{1}{V_s}\sum_{i; M_i\ge M_{\rm th}}\hat N_i
p_i(k)=\hat{P}^{2h}(k)+\frac{1}{V_s}\sum_{i; M_i<M_{\rm
th}}\hat N_i\hat{p}^{1h}(k)
+\frac{1}{V_s}\sum_{i;M_i\ge M_{\rm
th}}\hat N_i\left[\hat{p}^{1h}_i(k)-p^{1h}_i(k)\right].
\label{eq:est_dpk2}
\end{equation}
The ensemble average of the estimator (\ref{eq:est_dpk2}) yields
\begin{eqnarray}
\ave{\widehat{\Delta P}(k)}&=&
P^{2h}(k)+\int_0^{M_{\rm th}}\!\!\rmd M~ \frac{\rmd n}{\rmd M}p_M^{1h}(k)\nonumber\\
&\equiv & P^{2h}(k)+P^{1h}_{M<M_{\rm th}}(k),
\end{eqnarray}
where we have introduced the notation defined as $P^{1h}_{M<M_{\rm
th}}\equiv \int_0^{M_{\rm th}}\rmd M~(\rmd n/\rmd M)p^{1h}_M(k)$, the
1-halo term contribution arising from halos with masses $M<M_{\rm
th}$. Thus the estimator reduces the 1-halo term arising form massive
halos with $M>M_{\rm th}$.

Likewise, we can compute the covariance matrix of the Gaussianized power
spectrum estimator:
\begin{eqnarray}
{\rm Cov}[\widehat{\Delta P}(k),\widehat{\Delta P}(k')]_{M_{\rm th}}
&=&\frac{2}{ N_{\rm mode}(k)}
\delta^K_{kk'}
\left[P^{2h}(k)+P^{1h}_{M<M_{\rm th}}(k)\right]^2
+\frac{1}{V_s}
\left[\bar{T}(k,k')
-\bar{T}^{1h}(k,k'; M>M_{\rm th})
\right]
\nonumber\\
&&\hspace{2em}
+\left[
\int^{M_{\rm th}}_0\!\rmd M\frac{\rmd n}{\rmd M}b(M)p^{1h}_M(k)
\right]
\left[
\int^{M_{\rm th}}_0\!\rmd M'\frac{\rmd n}{\rmd M'}b(M')p^{1h}_{M'}(k')
\right]\sigma_m^2(V_s),
\end{eqnarray}
where $T^{1h}(k,k; M>M_{\rm th})$ is the 1-halo term of matter
trispectrum containing only the contributions from massive halos with
$M>M_{\rm th}$.  The estimator does suppress the non-Gaussian error
contributions arising from massive halos with masses $M>M_{\rm th}$.
Then the question we want to address is whether the modified power
spectrum estimator can recover the information content.

To be comprehensive, 
the cross-covariance between the number counts and the power spectrum
is
\begin{eqnarray}
{\rm Cov}[\hat N_i, \widehat{\Delta P(k)}]&=&
b_{i}\bar{N}_{i} \sigma_m^2\int^{M_{\rm th}}_0\!\!\rmd M\frac{\rmd
n}{\rmd M} b(M)p_M^{1h}(k)
\nonumber\\
&&+\bar{N}_i\sigma_m^2
\int\!\rmd M\rmd M'\frac{\rmd n}{\rmd M}\frac{\rmd n}{\rmd M'}\left[
b(M_i)b(M)+b(M_i)b(M')+b(M)b(M')
\right]p^{2h}_{MM'}(k).
\end{eqnarray}
The power spectrum estimator (Eq.~\ref{eq:est_dpk}) suppresses the 1-halo
term contribution of the cross-covariance arising form massive halos
with $M>M_{\rm th}$.

\section{Application to weak lensing power spectrum}
\label{sec:wl}

Since the weak lensing power spectrum is a projection of the three dimensional
power spectrum, we can extend the 
covariance
calculations and the Gaussianization
methodology of Section~\ref{sec:formulation} to the weak lensing observables.

\subsection{Angular number counts of halos, weak lensing power spectrum
  and their covariance matrices}
  
There are a number of potential methods of obtaining a mass-limited halo
sample.  Perhaps, the most attractive approach is to simultaneously
carry out a CMB survey and a weak lensing survey for the same
region of the sky.  In the next few
years, the HSC and the new-generation high-angular resolution,
high-sensitivity CMB experiment, ACTPol \citep{ACTPol:10},  
will survey overlapping regions of the sky as will
the DES and the SPTPol \citep{SPTPol:12}. 
The weak lensing surveys will calibrate the SZ flux-mass
relation and the SZ surveys will provide a mass-selected sample of halos
for the joint analysis envisioned in this paper. Further, if the imaging
survey has overlapping footprints with a wide-area spectroscopic survey,
the spectroscopic data can determine redshifts of the identified
clusters from the spectroscopic redshifts of member galaxies, such as
BCGs, 
and/or can
improve an identification of massive clusters from a concentration of
the spectroscopic member galaxies in the small spatial region
\citep{ReidSpergel:09,Hikageetal:12,Masakietal:13}. This is indeed the
case for combinations of the HSC survey with the BOSS or PFS surveys,
and the space-based Euclid and WFIRST projects. Soon
the all-sky eROSITA
survey, scheduled to be launched in
2015\footnote{\url{http://www.mpe.mpg.de/eROSITA}},
can be used to improve
the completeness/purity of the cluster catalog. 

In a full analysis, we would need to include
the scatter in the mass observable relation; however, for this paper, we will simplify
the presentation by assuming that we can directly count 
the number of halos in the
mass range of $[M_i, M_i+\Delta M]$ and over the entire redshift range
$0<z<z_{\rm max}$, in a light-cone volume with solid
angle $\Omega_s$:
\begin{equation}
\bar{N}^{2D}_{i}\equiv \Omega_s \int_0^{\chi_H}\!\!\rmd\chi~ \chi^2
 \left.\frac{\rmd n}{\rmd M}\right|_{M_i}\Delta M.
\label{eq:n_2D}
\end{equation}
%
The estimator for the angular number counts of
halos is given as
\begin{equation}
\hat{N}^{2D}_{i}=\sum_{\tilde{b}=1}\hat{N}_{i(\tilde{b})},
\end{equation}
where $\hat{N}_{i(b)}$ is the {\em observed} 
number counts of halos in the mass
range $[M_i,M_i+\Delta M]$ and in the redshift range
$[\chi_b,\chi_b+\Delta \chi]$, which has the comoving volume of 
$\Omega_s\chi^2\Delta
\chi$.
The ensemble average
 $\ave{\hat{N}_{i(b)}}\equiv \bar{N}_{i(b)}=\Omega_s
\chi_b^2\Delta \chi (\rmd n/\rmd M_i) \Delta M$. 
Hereafter we sometimes use notation ``$b$'' or ``$b'$'' to
denote the $b$- or $b'$-th redshift bin (do not confuse with the halo
bias $b(M_i)$ or $b_i$). 

Similarly to the derivation used in Eqs.~(\ref{eq:n_cov1}) and
(\ref{eq:n_cov2}), 
the covariance matrix of the number counts (Eq.~\ref{eq:n_2D}) are computed as
\begin{eqnarray}
{\rm Cov}\left[
\hat{N}^{2D}_{i},\hat{N}^{2D}_{j}
\right]&=& \bar{N}^{2D}_{i}\delta_{ij}^K+
\sum_{\tilde b}
\bar{N}_{i(\tilde b)}\bar{N}_{j(\tilde
b)}\sigma_m^2(\Omega_{s};\chi_{\tilde b})\nonumber\\
&&\hspace{-7em}=\bar{N}^{2D}_{i}\delta^K_{ij}+
\sum_{\tilde b=1}
\left[
\Omega_{\rm s}\chi_{\tilde b}^2\Delta \chi 
b(M_i)\left.\frac{\rmd n}{\rmd M}\right|_{M_i}\Delta M
\right]
\left[
\Omega_{\rm s}\chi_{\tilde b}^2\Delta \chi 
b(M_j)\left.\frac{\rmd n}{\rmd M}\right|_{M_j}\Delta M
\right]
\sigma_m^2(\Omega_{s}; \chi_{\tilde b})\nonumber\\
&&\hspace{-7em}\simeq 
\bar{N}^{2D}_{i}\delta^K_{ij}+
\Omega_s^2\int^{\chi_b}_0\!\!\rmd\chi~ 
\left[
\chi^2
b(M_i)\left.\frac{\rmd n}{\rmd M}\right|_{M_i}\Delta M
\right]
\left[
\chi^2
b(M_j)\left.\frac{\rmd n}{\rmd M}\right|_{M_j}\Delta M
\right]
\Delta \chi \sigma_m^2(\Omega_{s}; \chi)\nonumber\\
&&\hspace{-7em}\simeq
\bar{N}^{2D}_{i}\delta^K_{ij}+
\Omega_s^2\int^{\chi_b}_0\!\!\rmd\chi~ 
\left[
\chi^2
b(M_i)\left.\frac{\rmd n}{\rmd M}\right|_{M_i}\Delta M
\right]
\left[
\chi^2
b(M_j)\left.\frac{\rmd n}{\rmd M}\right|_{M_j}\Delta M
\right]
\int\!\!\frac{\rmd^2\bm{k}_\perp}{(2\pi)^2}
P^L_m(k_\perp;\chi)\left|
\tilde{W}_\perp(\bm{k}_\perp;\Omega_s)
\right|^2,
\label{eq:cov_n2d}
\end{eqnarray}
where 
$\sigma^2_m(\Omega_s;\chi_b)$ is the rms mass
fluctuations of the volume around the $b$-th redshift bin $\chi_b$, 
$V(\chi_b)=\Omega_s\chi_b^2\Delta \chi$,
and $\tilde{W}_\perp(\bm{k}_\perp;\Omega_s)$ is the Fourier
transform of the angular survey window function assuming the flat-sky
approximation.
In the third line on the r.h.s., we backed the summation to the
integration.
We have also assumed that the halo number counts of different redshift
bins are uncorrelated with each other; to be more precise, we used the
following ensemble average:  
\begin{equation}
\ave{N_{i(b)}N_{j(b')}}=\bar{N}_{i(b)}\bar{N}_{j(b')}
+\delta^K_{bb'}\delta^K_{ij}\bar{N}_{i(b)}+
\delta^K_{bb'}\bar{N}_{i(b)}\bar{N}_{j(b)}\left[
1+b_{i}b_{j}\sigma_m^2(\Omega_s; \chi_b)
\right],
\end{equation}
where $b_i\equiv b(M_i)$ and so on.
In the fourth line on the r.h.s., we used the
following calculation, based on the Limber's approximation \citep{Limber:54}:
\begin{eqnarray}
\Delta \chi \sigma^2_m(k;
 \Omega_s,\chi)&=&\Delta\chi\int\!\frac{\rmd^2\bm{k}_\perp}{(2\pi)^2}
\int^{\infty}_{-\infty}\!\frac{\rmd k_\parallel}{2\pi}P^L_m(k;\chi)\left|
\tilde{W}_\perp(\bm{k}_\perp;\Omega_s)
\right|^2 |W_\parallel(k_\parallel;\chi)|^2\nonumber\\
&\simeq & \Delta\chi\int\!\frac{\rmd^2\bm{k}_\perp}{(2\pi)^2}
P^L_m(k_\perp;\chi)\left|
\tilde{W}_\perp(\bm{k}_\perp;\Omega_s)
\right|^2 
\int^{\infty}_{-\infty}\frac{\rmd k_\parallel}{2\pi}
\left|
\tilde{W}_\parallel(k_\parallel)
\right|^2\nonumber\\
&=&\int\!\!\frac{\rmd^2\bm{k}_\perp}{(2\pi)^2}
P^L_m(k_\perp;\chi)\left|
\tilde{W}_\perp(\bm{k}_\perp;\Omega_s)
\right|^2,
\label{eq:sigm_dchi}
\end{eqnarray}
where $W_\parallel$ is the radial selection function, and we have
assumed in the second line on the r.h.s that only the density
fluctuations with Fourier modes perpendicular to the line-of-sight
direction contribute the rms of super-survey modes. For the integration
of the radial selection function, we have used the identity \citep[see
Eq.~8 in][]{TakadaHu:13}:
$\int\!d\chi W_\parallel(\chi)=
\Delta\chi\int\!d\chi
W_\parallel(\chi)^2=\Delta\chi\int\!\frac{dk_\parallel}{2\pi}
|\tilde{W}_\parallel(k_\parallel)|^2=1$.
We have assumed 
that the radial bin width is sufficiently large compared to the
wavenumber $k$ relevant for the power spectrum of interest. 
Eq.~(\ref{eq:cov_n2d}) matches
Eq.~(13) in \citet{TakadaBridle:07}.

The weak lensing angular power spectrum is a projection of the matter power spectrum.
Thus, we can use 
 the halo model approach and the Limber's approximation 
to represent  the weak lensing power spectrum as the  sum of the 1- and 2-halo terms:
\begin{equation}
 C_l=C_l^{1h}+C_l^{2h},
\end{equation}
where $C_l^{1h}$ and $C_{l}^{2h}$ are the 1- and 2-halo terms defined as
\begin{eqnarray}
C_l^{1h }&=&\int\!\rmd\chi\frac{\rmd^2V}{\rmd\chi \rmd\Omega}
\int\!\rmd M\frac{\rmd n}{\rmd M} \left|\kappa_M(l; \chi)\right|^2
=\frac{1}{\Omega_s}\int\!\rmd\chi
\left[
\Omega_s \chi^2
\int\!\rmd M\frac{\rmd n}{\rmd M}\right] \left|\kappa_M(l; \chi)\right|^2,\\
C_l^{2h}&=&\int\!\rmd\chi W^{\rm GL}(\chi)^2\chi^{-2}
 P^{2h}\!\left(k=\frac{l}{\chi}; \chi\right)
\nonumber\\
&=&
\frac{1}{\Omega_{\rm s}^2}\int\!\rmd\chi~ \chi^{-6}\left[
\Omega_s \chi^2 \int\!\!\rmd M\frac{\rmd n}{\rmd M}
\right]\left[\Omega_s \chi^2 \int\!\rmd M'\frac{\rmd n}{\rmd M'}
\right]\frac{M}{\bar\rho_m}\frac{M'}{\bar\rho_m}
W_{\rm GL}^2 P_{hh}(k; M,M'),
\end{eqnarray}
where $k=l/\chi$,
$\tilde{\kappa}_M(l;\chi)$ is the two-dimensional Fourier
transform of the projected NFW profile \citep[see Eq.~28 in][for the
definition]{OguriTakada:11}, and $W_{\rm GL}$ is the lensing efficiency
function \citep[see Eq.~19 in][]{OguriTakada:11} that has a dimension
of $[{\rm Mpc}^{-1}]$. From the above equation, 
we find that
the
estimators of 1- and 2-halo term lensing power spectra can be rewritten as functions of 
observed halo number counts:
\begin{eqnarray}
\hat{C}^{1h}_l &\equiv& \frac{1}{\Omega_s}\sum_b\sum_{i}\hat{N}_{i(b)}
\hat{C}^{1h}_i(l;\chi_a),\nonumber\\
\hat{C}^{2h}_l &\equiv & \frac{1}{\Omega_s^2\Delta \chi}
\sum_b\sum_{i,j}\hat{N}_{i(b)} \hat{N}_{j(b)}\chi^{-6}_b
\hat{P}^{2h}_{ij}\!\!\left(k=\frac{l}{\chi_b}; \chi_b\right)
\end{eqnarray}
where $\hat{C}_i^{1h}(l)$ is the estimator arising from the mass density
field inside halos of mass $M_i$, and 
 $\hat{P}^{2h}_{ij}(k)$ arises from the mass density field at
 large scales that governs the distribution between different halos of masses
 $M_i $ and $M_j$. 
 Note that
$\hat{P}^{2h}_{ij}$ is defined as
$\hat{P}^{2h}_{ij}\equiv (M_i/\bar\rho_m)(M_j/\bar\rho_m)
W_{\rm GL}^2\hat{P}_{ij}^{2h}$, and has a dimension
of $[({\rm Mpc})^7]$.

Using the similar derivation to Eq.~(\ref{eq:pscov}), the covariance
matrix of the lensing power spectrum is found to be
\begin{eqnarray}
{\rm Cov}[\hat{C}_l,\hat{C}_{l'}]&=&\frac{2}{N_{\rm
 mode}(l)}C_l^2\delta^K_{ll'}
+\frac{1}{\Omega_s}\bar{T}(l,l')\nonumber\\
&&\hspace{-2em}+\int\!\!\rmd\chi~ \left[
\chi^2\int\!\rmd M\frac{\rmd n}{\rmd M}b(M)\left|\kappa_M(l;\chi)\right|^2
\right]
\left[
\chi^2\int\!\rmd M'\frac{\rmd n}{\rmd M'}b(M')\left|\kappa_{M'}(l';\chi)\right|^2
\right]\int\!\!\frac{\rmd^2\bm{k}_\perp}{(2\pi)^2}P^L_m(k_\perp)
\left|\tilde{W}_\perp(k_\perp;\Omega_s)\right|^2,
\nonumber\\
\label{eq:clcov}
\end{eqnarray}
where $N_{\rm mode}(l)\equiv l\Delta l \Omega_s/(2\pi)$ and $\Delta l$ is the
bin width.
Again the above equation reproduces Eq.~(18) in \cite{Satoetal:09}
\citep[see also Eq. 14 in][]{Kayoetal:13}. In reality an accuracy of the
lensing power spectrum measurement is affected by intrinsic shape
noise. Assuming random shape orientations in between different galaxies,
the measured lensing power spectrum is contaminated by the shape noise
as
\begin{equation}
 C_l^{\rm obs}=C_l + \frac{\sigma_\epsilon^2}{\bar{n}_g},
\end{equation}
where $\sigma_\epsilon$ is the rms of intrinsic ellipticity per
component and $\bar{n}_g$ is the mean number density of source
galaxies. By replacing $C_l$ with $C_l^{\rm obs}$ in
Eq.~(\ref{eq:clcov}), we can take into account the shape noise
contamination to the covariance matrix. We assume
$\sigma_{\epsilon}=0.22$ as for the fiducial value. 

Similarly, the cross-covariance between the angular number counts and
the lensing power spectrum is given as
\begin{eqnarray}
{\rm Cov}\left[\hat{N}^{2D}_{i},\hat{C}_l\right]&=&
\frac{1}{\Omega_s}\int_0^{\chi_b}\!\!\rmd\chi~\Omega_s \chi^2\frac{\rmd
n}{\rmd M_i}\Delta M\left|
\kappa_{M_i}(l;\chi)
\right|^2\nonumber\\
&&\hspace{-6em}+\int_0^{\chi_b}\!\rmd\chi~
 \Omega_s\chi^2b(M_i)\frac{\rmd n}{\rmd M_i}\Delta
M\left[\chi^2\int\!\rmd M\frac{\rmd n}{\rmd
  M}b(M)\left|\kappa_M(l;\chi)\right|^2\right]
\int\!\!\frac{\rmd^2\bm{k}_\perp}{(2\pi)^2}P^L_m(k_\perp)
\left|\tilde{W}_\perp(k_\perp;\Omega_s)\right|^2
\nonumber\\
&&\hspace{-6em}
+\int_0^{\chi_b}\!\rmd\chi~ \Omega_s\chi^2\frac{\rmd n}{\rmd M_i}\Delta
M
\chi^4 \int\!\rmd M\rmd M'~\frac{\rmd n}{\rmd M}\frac{\rmd n}{\rmd M'}
\left[2b(M_i)b(M)+b(M)b(M')\right]\nonumber\\
&&\times
\chi^{-6}W_{\rm GL}^2
\frac{M}{\bar\rho_0}
\frac{M'}{\bar\rho_0}
P_{hh}\!\!\left(k=\frac{l}{\chi};\chi, M,M'\right)
\int\!\!\frac{\rmd^2\bm{k}_\perp}{(2\pi)^2}P^L_m(k_\perp)
\left|\tilde{W}_\perp(k_\perp;\Omega_s)\right|^2.
\end{eqnarray}
Note that the shape noise does not contaminate to the cross-covariance.

\subsection{Joint likelihood function of the angular halo number counts
  and the weak lensing power spectrum}

Having derived all the covariance matrices of the halo number counts
and the weak lensing power spectrum as well as their cross-covariance, we
can advocate the joint likelihood
function for the two
observables. 
Assuming that the two observables obey a multivariate
Gaussian distribution, we can derive the joint likelihood function as
\begin{equation}
{\cal L}(\hat{N}_{i}^{2D},\hat{C}_l)\propto 
\exp\left[-\frac{\chi^2}{2}\right]\equiv 
\exp\left[-\frac{1}{2}\bm{D}^{T}\bm{C}^{-1}\bm{D}
\right],
\label{eq:joint_n-wl}
\end{equation}
where the data vector $\bm{D}$ and the covariance matrix $\bm{C}$ are
defined as 
\begin{eqnarray}
\bm{D}&\equiv& \left(
\begin{array}{cc}
\hat{N}_{i}^{2D} - \bar{N}_{i}^{2D}
& 
\hat{C}_l -\bar{C}_l
\end{array}
\right),
\nonumber\\
\bm{C}&\equiv& 
\left(
\begin{array}{cc}
{\rm Cov}[\hat{N}_{i}^{2D},\hat{N}_{j}^{2D}] & 
{\rm Cov}[\hat{N}_{i}^{2D},\hat{C}_l] \\
{\rm Cov}[\hat{C}_l,\hat{N}_{j}^{2D}] &
{\rm Cov}[\hat{C}_l,\hat{C}_{l'}]
\\
\end{array}
\right).
\label{eq:cov_n-wl}
\end{eqnarray}
Here $\bm{C}^{-1}$ is the inverse of
the covariance matrices. Note that the products of the data vector and
the inverse of the covariance matrix 
run over
all the halo mass
bins 
as well as the multipole bins.

We used the above equation to compute the joint likelihood function in
Fig.~\ref{fig:dn-dcl-model}. The 68 or 95\% percentile of the
distribution for the two observables is obtained from the range
satisfying $\chi^2\ge 2.3$ or $6.17$, respectively.  We again
notice that the halo model (Eq.~\ref{eq:joint_n-wl}) well
reproduces the simulation results, giving a justification of our method
and the multivariate Gaussian assumption at the angular scales.

\subsection{Information content of
the Gaussianized weak lensing power spectrum}

According to the discussion in Section~\ref{sec:dp}, we can define an
estimator of the power spectrum suppressing the 1-halo term
contribution, assuming that all halos with masses greater than a certain
mass threshold $M_{\rm th}$ in the surveyed light-cone volume
 are identified from a
given survey volume:
\begin{equation}
\left.\widehat{\Delta C_l}\right|_{M_{\rm th}}=\hat{C}_l -\frac{1}{\Omega_s}\sum_b \sum_{i;
 M>M_{\rm th}} 
\hat{N}_{i(b)}{C}^{1h}_i(l;\chi_b),
\end{equation}
where $\hat N_{i(b)}$ is the observed counts of halos with $M\ge M_{\rm
th}$ and in the redshift range of $[\chi_b,\chi_b+\Delta\chi]$, and
$C^{1h}_i(l; \chi_b)$ is the theory temperate of the 1-halo term power
spectrum for halos with $M_i$ and at redshift $\chi_b$. 
The theoretical template can be estimated from stacked lensing
of the sampled halos, e.g., using the method in \cite{OguriTakada:11}. 

The ensemble average of the estimator and the covariance are 
\begin{eqnarray}
\ave{\left. \widehat{\Delta C_l}\right|_{M_{\rm th}}}&=&
C_l^{2h}+\left.C_l^{1h}\right|_{M<M_{\rm th}}\equiv 
C_l^{2h}+\int\!\rmd\chi~\frac{\rmd^2V}{\rmd\chi \rmd\Omega}\int^{\rm M_{\rm
th}}_0\!\rmd M~\frac{\rmd n}{\rmd M}\left|\kappa_M(l;\chi)\right|^2,
\label{eq:cl_r1h_exp}
\end{eqnarray}
and
\begin{eqnarray}
{\rm Cov}\left[
\left.\widehat{\Delta C_l}\right|_{M_{\rm th}},
\left.\widehat{\Delta C_{l'}}\right|_{M_{\rm th}}
\right]&=&
\frac{2}{N_{\rm mode}(l)}
\left[
C^{2h}_l + \left. C_l^{1h}\right|_{M<M_{\rm th}}
+\frac{\sigma_\epsilon^2}{\bar{n}_g}\right]^2\delta^K_{ll'}
+\frac{1}{\Omega_s}
\left[
\bar{T}_\kappa(l,l')
-\bar{T}_\kappa^{1h}(l,l'; M>M_{\rm th})
\right]
\nonumber\\
&&\hspace{-12em}+
\int\!\!\rmd\chi~ \left[
\chi^2\int_0^{M_{\rm th}}\!\rmd M\frac{\rmd n}{\rmd M}b(M)
\left|\kappa_M(l;\chi)\right|^2
\right]
\left[
\chi^2\int_0^{M_{\rm th}}\!\rmd M'\frac{\rmd n}{\rmd M'}b(M')\left|\kappa_{M'}(l';\chi)\right|^2
\right]\int\!\!\frac{\rmd^2\bm{k}_\perp}{(2\pi)^2}P^L_m(k_\perp)
\left|\tilde{W}_\perp(k_\perp;\Omega_s)\right|^2,\nonumber\\
\label{eq:cl_r1h_cov}
\end{eqnarray}
where we have also included the intrinsic shape noise contribution. 
In the following, we assume 
a circular-shaped
survey geometry with area $\Omega_s=\pi\Theta_s^2$, yielding
$\tilde{W}_\perp(k;\Omega_{s})=2J_1(\chi k\Theta_s)/(\chi k\Theta_s)$.

The information content inherent in the power spectrum measurement for a
given survey is defined in \cite{Tegmarketal:97} \citep[see also][]{TakadaJain:09} as
\begin{equation}
\left(\frac{S}{N}\right)^2\equiv \sum_{l_i,l_j<l_{\rm max}}
\left.{\Delta C_{l_i}}\right|_{M_{\rm th}}
[{\bf{C}^{-1}}]_{l_il_j}
\left.{\Delta C_{l_j}}\right|_{M_{\rm th}},
\label{eq:sn}
\end{equation}
where the summation runs over multipole bins up to a given maximum
multipole $l_{\rm max}$, $\left.{\Delta C_l}\right|_{M_{\rm th}}$ is the
expectation value of the power spectrum (Eq.~\ref{eq:cl_r1h_exp}), and
$\bm{C}^{-1}$ denotes the inverse of the covariance matrix
(Eq.~\ref{eq:cl_r1h_cov}). The inverse of $S/N$ is equivalent to a
fractional error of measuring the power spectrum amplitude when using
the information up to the maximum multipole $l_{\rm max}$, 
assuming that the shape of
the power spectrum is perfectly known. For a Gaussian field, Eq. (\ref{eq:sn})
predicts
 $S/N\propto l_{\rm max}$. 

\subsection{Results}

\begin{figure}
\centering
\includegraphics[width=0.5\textwidth,angle=-90]{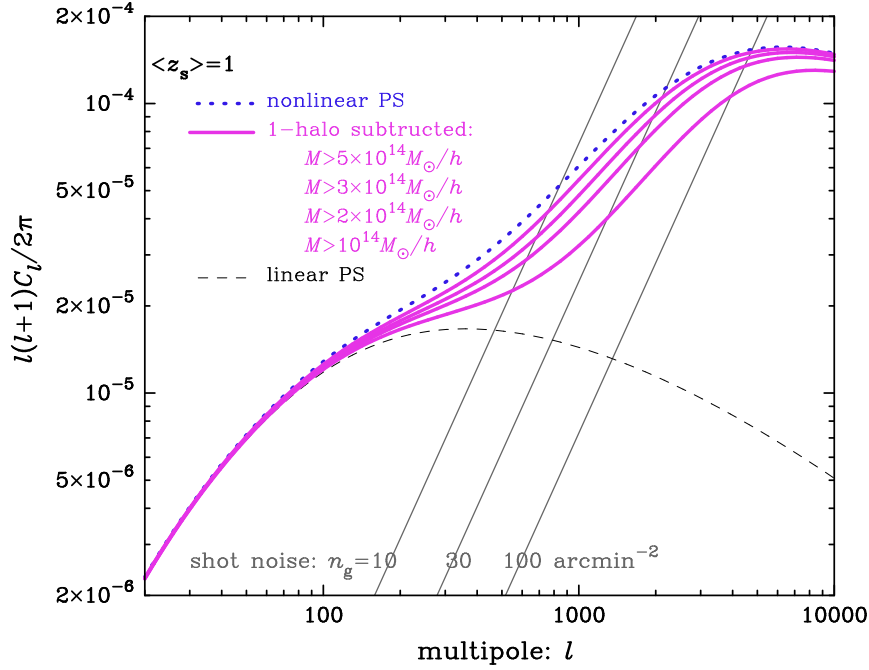}
\caption{Weak lensing power spectrum for source galaxies whose mean
 redshift $\langle z_s\rangle=1$. The dashed curve is the spectrum
 obtained by projecting the linear matter power spectrum weighted with
 the lensing efficiency function, while the top dotted curve is the
 result for the nonlinear matter power spectrum. The solid curves are
 the spectra where the 1-halo term contribution arising from massive
 halos with $M>1, 2, 3 $ or $5~ [10^{14}M_\odot/h]$ is
 subtracted, respectively. For comparison, the thin solid curves are the
 relative shape noise contamination for the number densities of
 $\bar{n}_g=10$, 30 and 100~arcmin$^{-2}$, respectively, 
where we assumed
 $\sigma_\epsilon=0.22$ for the rms intrinsic
 ellipticities.
\label{fig:cl}}
\end{figure}

\begin{figure}
\centering
\includegraphics[width=0.5\textwidth,angle=-90]{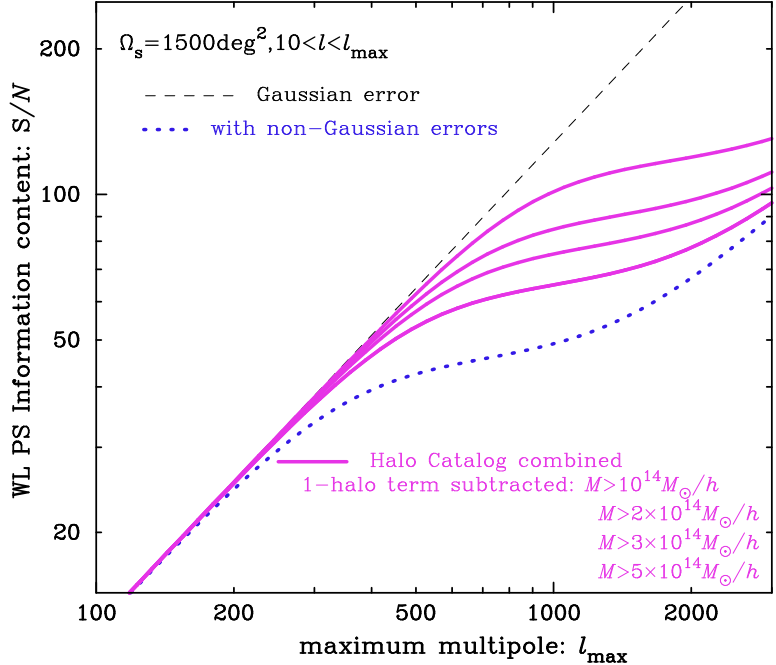}
\caption{Cumulative signal-to-noise ($S/N$) ratio (Eq.~\ref{eq:sn}) for
the weak lensing power spectrum measurement as a function of maximum
multipole $l_{\rm max}$, expected for a wide-area survey that is
characterized by $\Omega_{\rm s}=1500$~sq. degrees and $\langle
z_s\rangle=1$ for survey area and the mean redshift of source galaxies,
respectively. Note that we did not include the shape noise
contamination. The top dashed line is the $S/N$ for a Gaussian field,
which has a scaling of $S/N|_{\rm Gaussian}\propto l_{\rm max}$. The
bottom dotted curve is the $S/N$ computed by using the full non-Gaussian
error covariance including the halo sampling variance contribution. The
solid curves are the $S/N$ values for a {\em Gaussianized} weak lensing
field, where the 1-halo term contribution for halos with masses greater
than a given mass threshold, as indicated by the legend, is subtracted
assuming that such massive halos are identified in the survey region.
The Gaussianization method using massive halos with $M\simgt
10^{14}M_\odot/h$ recovers the information content by up to a factor of
a few, especially over $300 \simlt l_{\rm max}\simlt 3000$.
\label{fig:sn_wosn}}
\end{figure}

\begin{figure}
\centering
\includegraphics[width=0.4\textwidth,angle=-90]{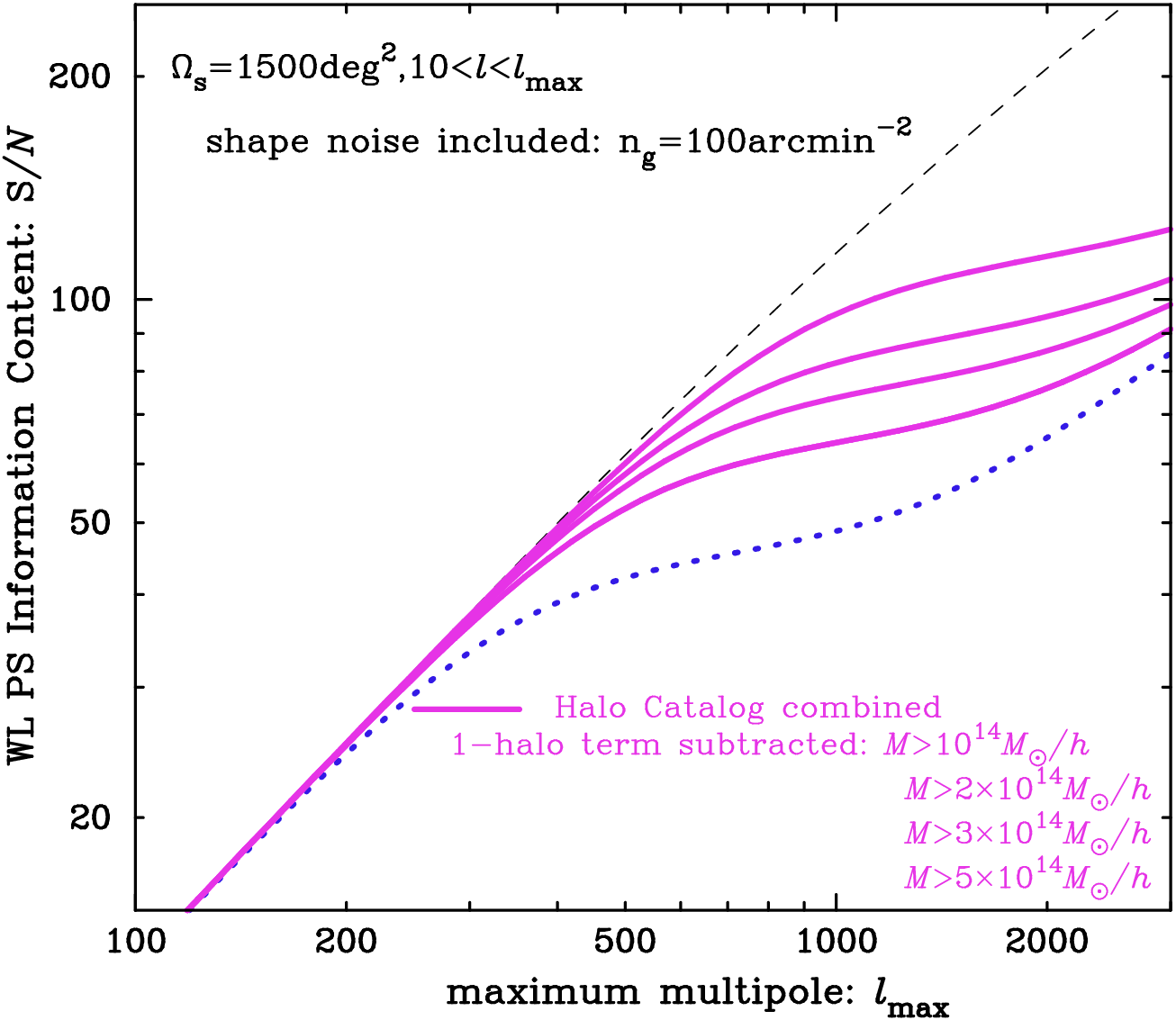}
\includegraphics[width=0.4\textwidth,angle=-90]{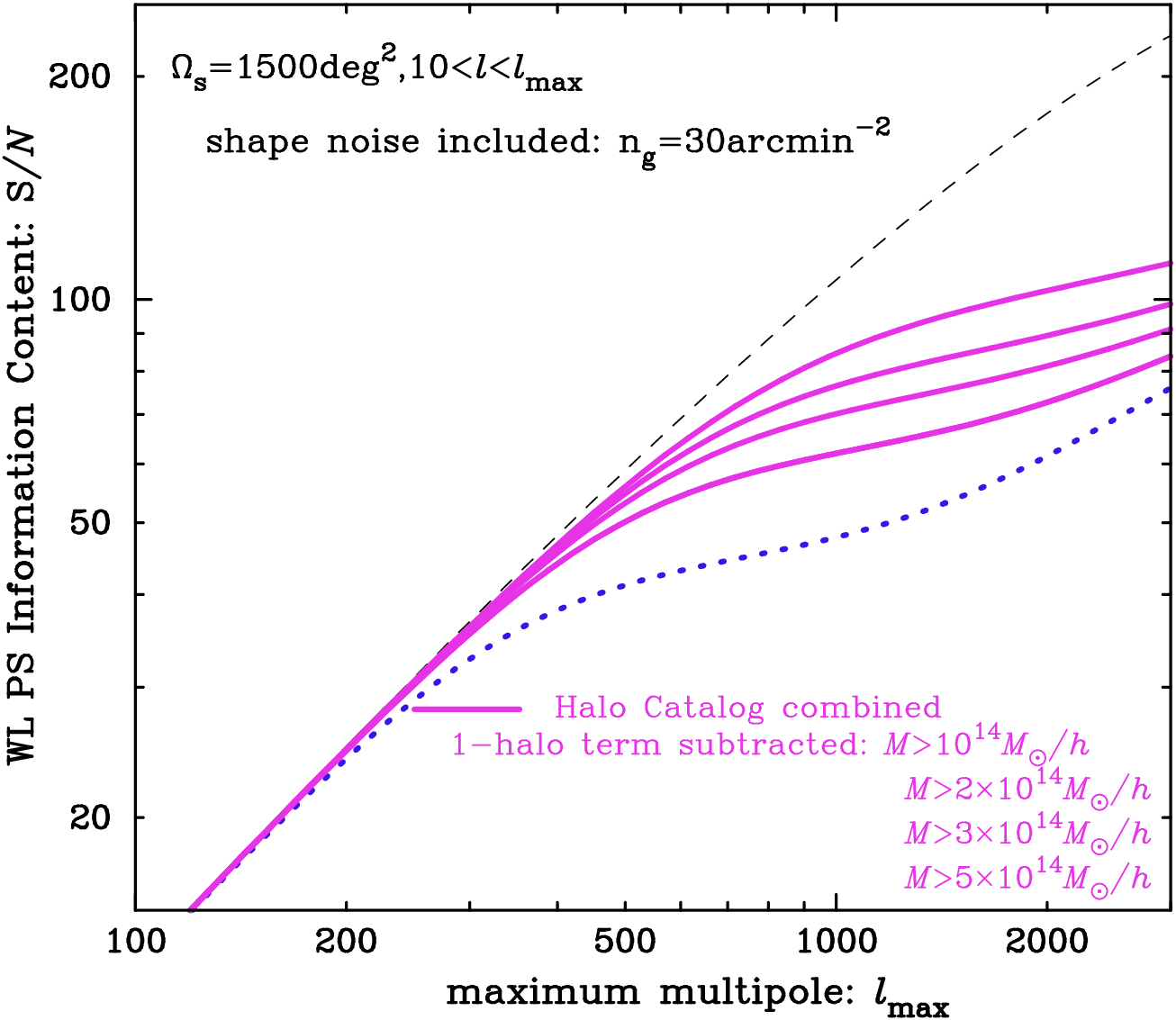}
\caption{Similar to the previous figure, but the shape noise
 contamination is included in the power spectrum covariance
 calculation. We here considered $\bar{n}_g=100$  or
 30~arcmin$^{-2}$ 
for the number density of galaxies in the left or right panels, 
respectively, which roughly correspond to the number densities for
the  WFIRST and LSST-type surveys, respectively. 
Other survey parameters are kept fixed
to the fiducial values, $\Omega_{s}=1500$
 deg$^2$ and $\langle z_s\rangle=1$. Note that 
the top curve for a Gaussian field includes 
the shot noise contamination in the covariance calculation.
\label{fig:sn_ng100}}
\end{figure}

\begin{figure}
\centering
\includegraphics[width=0.4\textwidth,angle=-90]{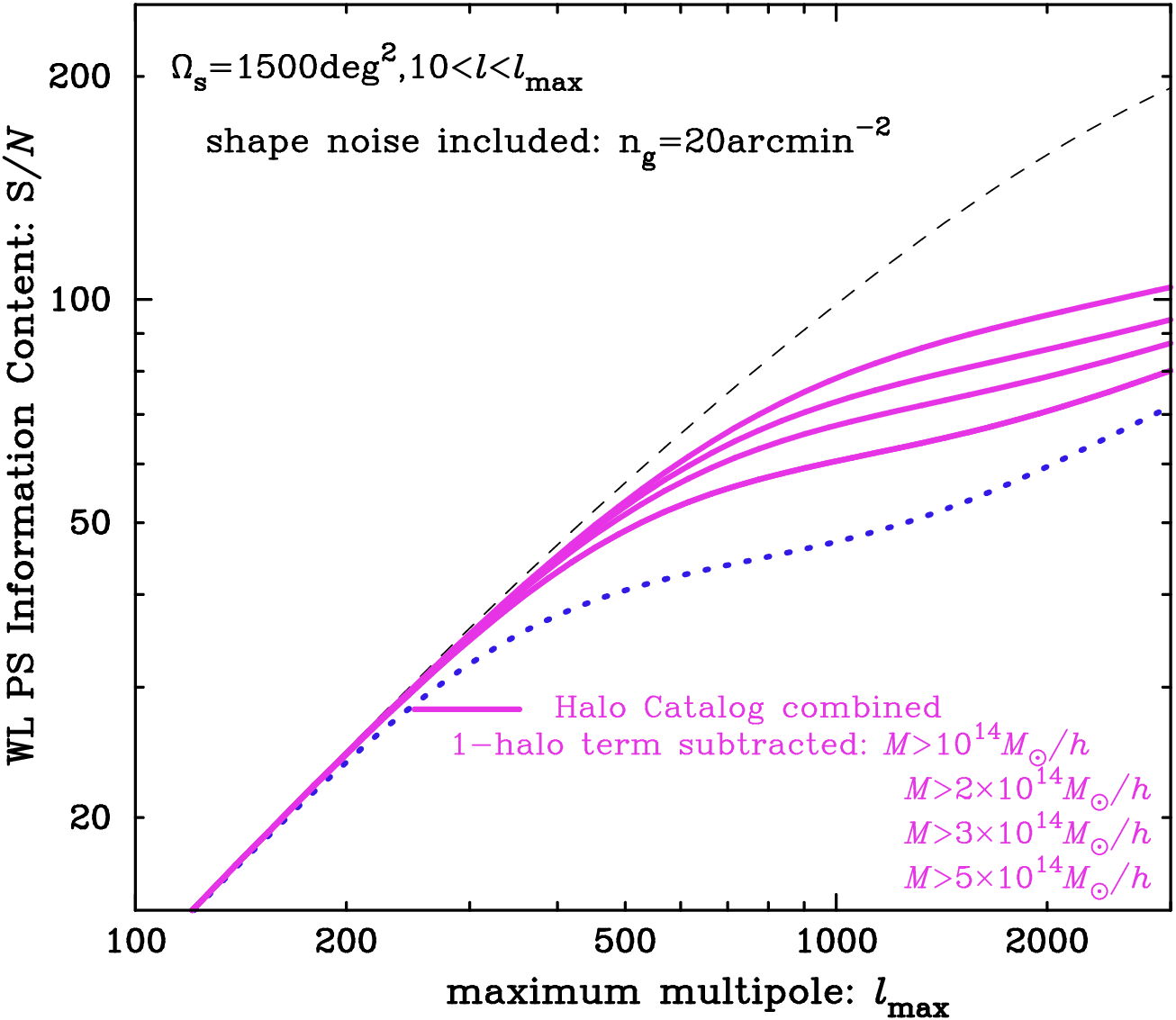}
\includegraphics[width=0.4\textwidth,angle=-90]{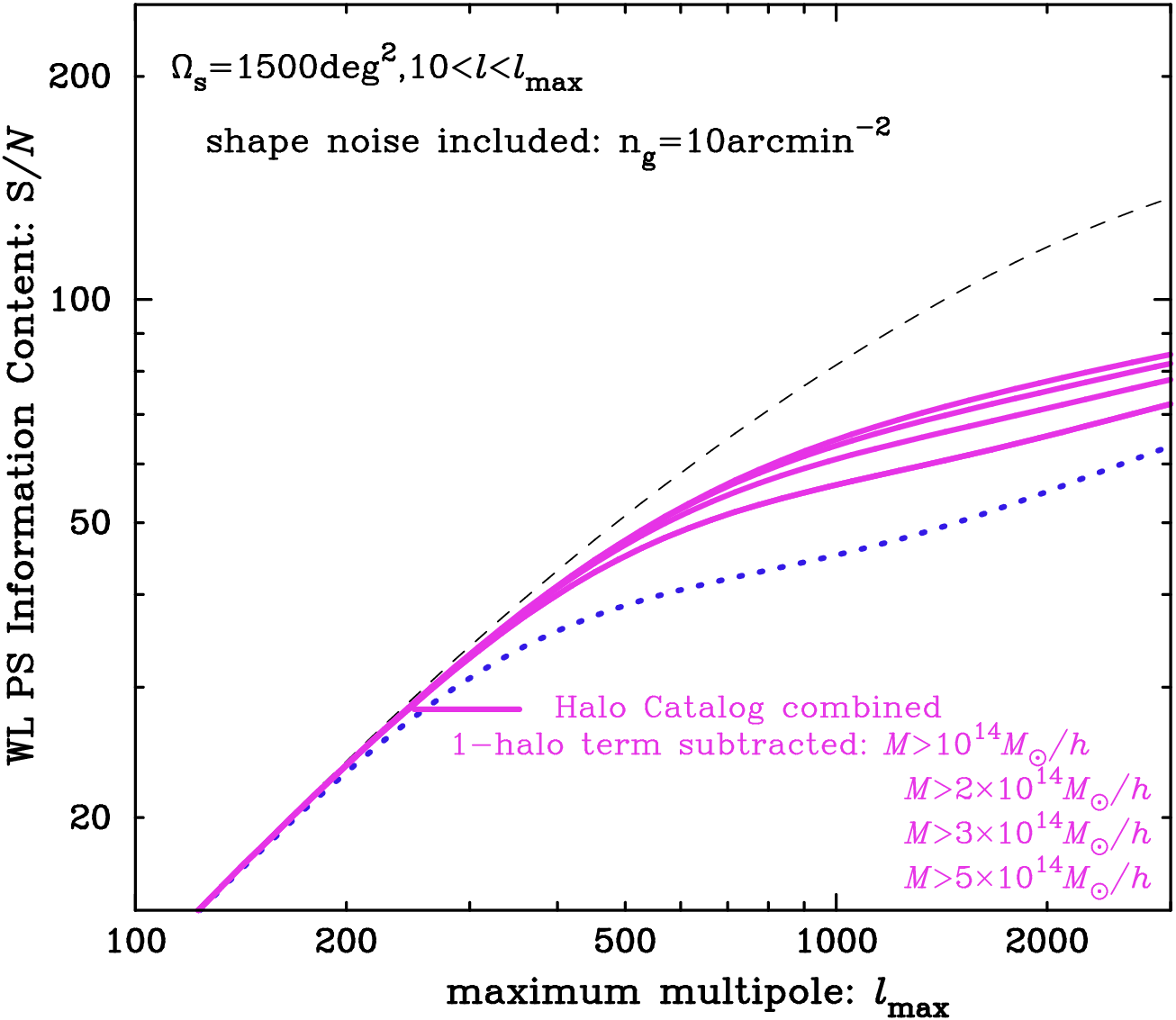}
\caption{Similar to the previous figure, but we assumed 
$\bar{n}_g=20$ or 10~arcmin$^{-2}$ in the left or right panels,
which roughly correspond to the Subaru HSC- or DES/Euclid/KiDS-type 
surveys, respectively.
\label{fig:sn_ng20}}
\end{figure}
To estimate an expected performance of the Gaussianized weak lensing field for
an upcoming imaging survey, we need to assume the fiducial cosmological
model and the survey parameters. For the fiducial cosmological model, we
assume a $\Lambda$CDM model that is consistent with the WMAP 7-year
result in \citet{Komatsuetal:10}. We use the same model ingredients in
\cite{OguriTakada:11} to compute the halo model predictions. As for
survey parameters, we employ the parameters that resemble the planned
Subaru Hyper Suprime-Cam (HSC) survey: $\langle z_s \rangle=1$,
$\bar{n}_g=20~$arcmin$^{-2}$, $\sigma_\epsilon=0.22$ and $\Omega_{\rm
s}=1500$~deg$^2$ for the mean redshift of galaxies, the mean number
density, the rms intrinsic ellipticities, and the survey area,
respectively. For the redshift distribution of imaging galaxies, we
employ Eq.~(17) in \citet{OguriTakada:11}, where we set the parameter
$z_0=1/3$ so as to have $\langle z_s\rangle =3z_0=1$. We will also study
how the results are changed by varying the shot noise contamination.

The solid curves in Fig.~\ref{fig:cl} show the power spectra when the
1-halo term contribution arising from massive halos with $M\ge 1, 2, 3$
or 5 $[10^{14}M_\odot/h]$ is subtracted, respectively.
The subtraction reduces the power spectrum amplitudes at high
multipole. The 
lower
mass cuts produce the greatest suppression of power.

Fig.~\ref{fig:sn_wosn} shows the expected cumulative signal-to-noise
ratio ($S/N$) for a Subaru HSC-type survey when implementing the
Gaussianization method of weak lensing power spectrum, combined with the
number counts of massive halos. Here the ``cumulative'' $S/N$ is
obtained by integrating the signal-to-noise ratio of the weak lensing
power spectrum over angular scales from $l=10$ up to a given maximum
multipole $l_{\rm max}$ as denoted in the $x$-axis. Note that, for the
results in this plot, we did not include the shape noise contamination
in order to study a best-available improvement of the Gaussianization
method.  The top dashed line is the $S/N$ for a two-dimensional Gaussian
field, which gives the maximum $S/N$ value as the weak lensing field or
the underlying matter distribution originates from the initial Gaussian
field as in the CMB field. The bottom dotted curve is the $S/N$ value
when using the full power spectrum covariance including the non-Gaussian
errors, where the non-Gaussian errors due to super-survey modes gives a
dominant contribution at $l_{\rm max}\simgt $a few 100. The dotted curve
is similar to Fig.~9 in \citet{Satoetal:09} \citep[see also Fig.~10
in][]{Kayoetal:13}.

The solid curves show the results when implementing the Gaussianization
method. More precisely, the curves are the results for the Gaussianized
power spectra, where the 1-halo term power spectrum arising from massive
halos with $M\ge 1, 2,3,$ or $5~ [10^{14}~M_\odot/h]$, respectively, is
subtracted from the total power, assuming that the massive halos of each
mass range are identified in the survey region.  The 1-halo term
subtraction significantly increases the information content.  With
decreasing the mass threshold, it recovers the information content,
almost the full information as does in a Gaussian field, up to higher
$l_{\rm max}$. 
To be more precise, the Gaussianization method recovers about 80
or 60 per cent at $l_{\rm max}=1000$ for $M_{\rm th}=1$ or
$3\times 10^{14}~M_\odot/h$, while it recovers about 50 or 30 per cent at $l_{\rm
max}=2000$, respectively.
Compared to the
$S/N$ value without the Gaussianization method, the improvement is up to
a factor of 2 or 1.4 for $M_{\rm th}=1$ or $3\times 10^{14}~M_\odot/h$
for the range of $l_{\rm max}=1000$--2000, 
which is equivalent to a factor 4 or 2 larger survey, respectively. The
improvement means that adding the abundance of massive halos to the
power spectrum measurement can correct for the super-sample covariance
contamination, because the super-sample covariance is a dominant source
to cause a saturation in the information content at $l_{\rm max}\simgt
500$ \citep[Fig.~9 in][]{Satoetal:09}. Even for the higher $l_{\rm.
max}$ such as $l_{\rm max}\simgt 2000$, where less massive halos of
$M\simlt 10^{14}M_\odot/h$ becomes important in the 1-halo term
contribution, the figure still displays a significant improvement, by up
to a factor of 2. Upcoming imaging surveys are aimed at constraining
cosmological parameters from the lensing power spectrum information up
to $l_{\rm max}=1000$--2000, beyond which 
complex baryonic physics in the nonlinear clustering 
can be important. Our results are very promising in
a sense that the Gaussianization method allows for an efficient masking
of the mass distribution in such a highly-nonlinear region, the region inside
massive halos, when measuring the power spectrum.

Figs.~\ref{fig:sn_ng100} and \ref{fig:sn_ng20} show the results when
including the shape noise contamination to the covariance, for different
number densities of source galaxies; $\bar{n}_g=10, 20, 30, $ and
100~arcmin$^{-2}$, respectively. The range includes the number densities
expected for the planned weak lens surveys; $70$ for WFIRST, 30 for
LSST/Euclid, 20 for HSC and 10~arcmin$^{-2}$ for DES/KiDS, respectively.
Note that other survey parameters (area and the mean redshift) are kept
fixed to their fiducial values as in Fig.~\ref{fig:sn_wosn}. The
relative improvement in the $S/N$ values with and without the
Gaussianization method does not largely change for different survey
areas.  The figures show that a survey having a higher number density
can have a greater benefit from the Gaussianization method as there is
more information on small scales that can be recovered.  To be
more precise, the Gaussianization method for a Subaru HSC-type survey
recovers about 90 or 70 per cent information of the Gaussian plus shape
nose case at $l_{\rm max}=1000$ for $M_{\rm th}=1$ or $3\times
10^{14}~M_\odot/h$, while it recovers about 75 or 55 per cent
information at $l_{\rm max}=2000$, respectively.  On the other hand, the
Gaussianization method gives about 1.6 improvement at $l_{\rm max}=1000$
-- 2000 compared to the results without the Gaussianization method for
$M_{\rm th}=10^{14}M_\odot/h$ (for a Subaru-type survey), while it gives
about 1.4 improvement for $M_{\rm th}=3\times 10^{14}M_\odot/h$. These
improvements are equivalent to a factor 2 -- 2.5 wider survey area. The
dependence of these results on survey area is very weak; therefore these
improvements hold for other surveys, as can also be found from Fig.~2 in
\citet{Kayoetal:13} or Fig.~1 in \citet{TakadaHu:13}.

\begin{figure}
\centering
\includegraphics[width=0.45\textwidth,angle=-90]{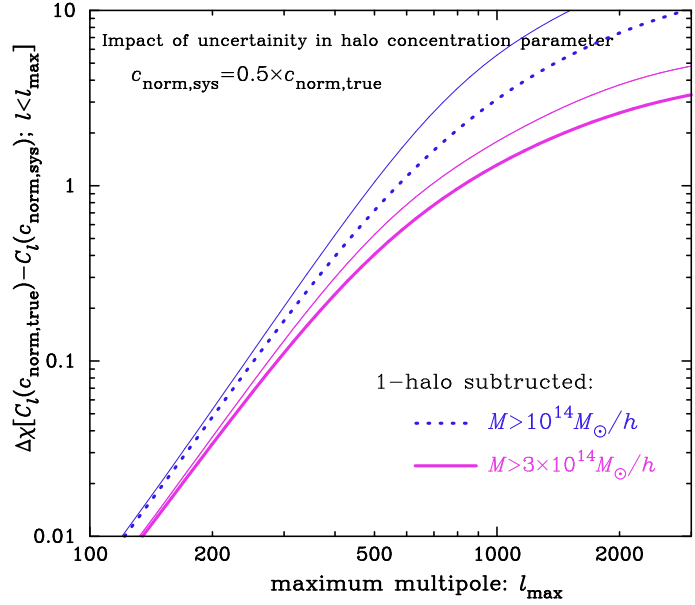}
\caption{Shown is how uncertainties in the 1-halo term power
 spectrum template, which is used for the 1-halo subtraction from the
 total power spectrum, affect the Gaussianization method. We model a
 misestimation of the 1-halo term template by including a bias in the
 normalization parameter of the halo concentration, $c(M,z)=c_0
 (1+z)^{-0.71}[M/(2\times 10^{12}M_\odot/h)]^{-0.081}$, with the
 fiducial value $c_0=7.85$ (see text for details). The plot shows the
 cumulative deviation between the true and misestimated power spectra up
 to a certain maximum multipole $l_{\rm max}$, when the average halo
 profile of halos with $M\ge M_{\rm th}$ is misestimated by an amount of
 $c_{0, {\rm sys}}=0.5c_{0, {\rm true}}$ (a factor 2 bias);
 $\Delta\chi^2=\sum_{l\le l_{\rm max}}\left[C_{l_i}(c_{0, {\rm true}})
 -C_{l_i}(c_{0,{\rm sys}})\right] {\bf{C}^{-1}}_{ij} \left[C_{l_j}(c_{0,
 {\rm true}}) -C_{l_j}(c_{0,{\rm sys}})\right] $ (see also text for the
 details). Note that $\sqrt{\Delta\chi^2}$ is plotted. The dotted and
 solid curves are the results for $M_{\rm th}=1$ and $3\times
 10^{14}~M_\odot/h$, respectively. The thick and thin respective curves
 are the results with and without the shape noise contamination for a
 Subaru HSC-type survey ($\bar{n}_g=20~$arcmin$^{-2}$ and $\Omega_{\rm
 s}=1500$ deg$^2$).  Although a factor 2 bias in the concentration
 parameter seems the worst case scenario, the cumulative deviation for
 $l_{\rm max}=10^3$ is only up to a few-$\sigma$ deviation for the
 Subaru-type survey.
 \label{fig:csys}}
\end{figure}
The Gaussianization method requires a knowledge of the average
mass profile for massive halos in order to subtract the inferred 1-halo
term from the total power spectrum.  The NFW profile seen in N-body
simulations is characterized by two parameters, the halo concentration
and halo mass. For lensing perspective, the halo mass is sensitive to
the area-weighted, integrated lensing signal up to the virial radius,
while the halo concentration needs to be estimated from the scale radius
which is the radius to divide the inner and outer profiles in the NFW
model.  We expect that stacked lensing of the massive halos can be used
to estimate the 1-halo profile
\citep{Okabeetal:10,OguriTakada:11,Okabeetal:13}, probably with the aid
of priors from N-body simulations. However, this estimate itself is
limited by statistical measurement uncertainties, suffers from
 degeneracies with cosmological
parameters, and can be affected by uncertainties in the astrophysical
effects such as baryonic effects on halo formation/structure
\citep[e.g.][]{Gnedinetal:04}. 

What is the required accuracy of the halo profile template? To
address this question, we study how variants in the halo profile affect
a performance of the Gaussianization method as follows.  We model the
halo profile variants by allowing a possible misestimation in the
normalization parameter of the halo mass and concentration relation,
$c(M,z)=c_0 (1+z)^{-0.71}[M/(2\times 10^{12}M_\odot/h)]^{-0.081}$, where
we employ $c_0=7.85$ as for the fiducial value following the N-body
simulation results in \citet{Duffyetal:08}.  This treatment is also
motivated by the study of \citet{Zentneretal:13}, where they showed that
the baryonic effects on the halo profile, which are indicated from
hydrodnynamical simulations, can be taken into account by including the
halo concentration parameters as nuisance parameters in weak lensing
cosmology.  \citet{OguriTakada:11} showed that, if the stacked lensing
measurement for a Subaru HSC-type survey is used to estimate the halo
profile parameters (the normalization, the mass slope and the
redshift-dependence slope) simultaneously with cosmological parameters,
the marginalized, fractional accuracy for the normalization parameter
$c_0$ is about 50 percent, i.e.  $|\sigma(c_0)|/c_0\simeq 0.5$, even
including possible miscentering effects of the halos. This accuracy is
considered as the worst case scenario, because in practice some priors
from N-body simulations can be used and/or the halo concentration
estimation can be improved by using detailed studies for representative
massive halo sample, e.g., based on the method combining strong and weak
lensing measurements \citep[e.g.][]{Broadhurstetal:05}.

Based on the above consideration, we assume that the halo mass
profile for massive halos, used for the 1-halo term subtraction, is
misestimated by an amount of factor 2, i.e. $c_{0,{\rm sys}}=0.5\times
c_{0,{\rm true}}$. Fig.~\ref{fig:csys} shows the cumulative deviation
between the true and model spectra for a Subaru HSC-type survey
($\bar{n}_g=20$~arcmin$^{-2}$ and $\Omega_s=1500$~deg$^2$):
\begin{equation}
\Delta\chi^2=\sum_{l\le l_{\rm max}}\left[C_{l_i}(c_{0, {\rm true}})
-C_{l_i}(c_{0,{\rm sys}})\right] {\bf{C}^{-1}}_{ij} \left[C_{l_j}(c_{0,
{\rm true}}) -C_{l_j}(c_{0,{\rm sys}})\right], 
\label{eq:dchi2}
\end{equation}
where $C_l(c_{0,{\rm true}})$ is the true spectrum with the true 1-halo
term being subtracted, $C_l(c_{0,{\rm sys}})$ is the model spectrum with
the biased 1-halo term being subtracted, and $\bm{C}^{-1}$ is the
inverse of the covariance matrix. The figure plots
$\Delta\chi=\sqrt{\Delta\chi^2}$; $\Delta\chi=1$ means the $\pm 1\sigma$
deviation between the true and model spectra.  The figure shows that,
even if we consider a factor 2 bias in the concentration parameter, the
cumulative deviations up to $l_{\rm max}=10^3$ are only up to a few
$\sigma$-deviation for $M_{\rm th}\ge 10^{14}M_\odot/h$ and a
Subaru-type survey. Recalling that the maximum multipole $l_{\rm
max}\sim 10^3$ corresponds to
$N_{\rm mod}\simeq \pi l_{\rm max}^2/[(2\pi)^2/\Omega_{s}]/2\sim 10^5$
for the total number of Fourier modes, where $(2\pi)/\sqrt{\Omega_s}$ is
the fundamental Fourier mode, only a few $\sigma$-deviation compared to
the huge data points is considered encouraging. This can be understood
as follows. The weak lensing signals up to $l_{\rm max}\sim 10^3$ are
not sensitive to the inner structure of halos or equivalently probe the
regime of $\tilde{u}_M(k)\simeq 1$ for the halo profile; the 1-halo term
up to $l_{\rm max}\sim 10^3$ is determined mainly by the abundance of
the massive halos (see Eq.~\ref{eq:ps_halomodel}).  The deviations
become increasingly larger for the larger maximum multipoles. Thus we
conclude that the requirement on the halo profile template needed for
the Gaussianization method is not so stringent; stacking lensing can
probably achieve the desired accuracy.

\section{Conclusion and Discussion}
\label{sec:conclusion}

In this paper, using the halo model approach and the likelihood function
of halo number counts, we have derived the joint likelihood function of
the halo number counts and the weak lensing
power spectrum.  The joint likelihood properly takes into account the
cross-correlation between the two observables  when they are measured from the same
survey region.  The cross-correlation in
the nonlinear regime is mainly due to the super-sample variance that
arises from the modes of length scales comparable with or greater than a
survey volume, which cannot therefore be directly observed. For instance, due to the super-survey sample variance, 
the weak lensing power
spectrum amplitudes around $l$ of a few thousands have a significant
correlation with the number counts of massive halos with $M\simgt
10^{14}~M_\odot/h$.
 We showed that our analytical model of the joint likelihood
function can well reproduce the distributions of halo number counts and
weak lensing power spectra seen from 1000 ray-tracing simulations (see
Fig.~\ref{fig:dn-dcl-model}). 

Given the strong correlation between the two observables, we have
proposed a method of combining the observed number counts of massive
halos with a measurement of matter or weak lensing power spectrum in the
same survey region, in order to suppress or correct for the super-sample
variance contamination -- the Gaussianization method of power spectrum
measurement (see Section~\ref{sec:wl}).  Massive halos with $M\simgt
10^{14}M_\odot$ are relatively easy to identify in a survey region,
e.g., from a concentration of member galaxies in the small spatial region
or $X$-ray and SZ observations if available. The Gaussianization can be
done by subtracting the 1-halo term power spectrum contribution,
weighted with the observed number counts of massive halos, from the
power spectrum. The method requires a theory template of the average
mass profile of the massive halos. In the paper, we can use the NFW profile  
 based on $N$-body simulations.  If we had survey data to use with this
 method, 
 we could 
use the stacked lensing method
\citep{OguriTakada:11,Okabeetal:10,Okabeetal:13} to directly estimate
the average mass profile around such massive halos from the data.  
This subtraction automatically corrects for the super-sample variance,
by using the observed number counts of massive halos that are affected
by super-survey modes. We showed that the weak lensing power spectrum
subtracting the 1-halo term can improve the information content, almost
recovering the full information content in a Gaussian field that should
have been in the initial density field as does the CMB field
(Fig.~\ref{fig:sn_wosn}).   If we can measure the number of halo 
 with $M\ge 1$ or $3\times
10^{14}M_\odot/h$, then the increase in the information content
can be up to a factor of 2 or 1.4 if the angular power spectrum is used
up to  $l_{\rm max}\simeq 2000$.  This is
equivalent to a factor 2 or 4 increase is  survey area. A survey having a larger
number density of galaxies, such as $\bar{n}_g=20$--100~arcmin$^{-2}$,
has a greater benefit from the Gaussianization method; the power
spectrum measurement is otherwise limited by the shape noise
contamination (Figs.~\ref{fig:sn_ng100} and \ref{fig:sn_ng20}).

The Gaussianization method suppressing the 1-halo term contamination in
the power spectrum measurement has an additional  practical advantage.
 Massive halos are a source of nonlinear clustering, and
 the matter distribution inside massive halos is in the deeply
nonlinear regime and is affected by complex baryonic physics that is
difficult to accurately model from first principles
\citep{HutererTakada:05,Sembolonietal:12,Zentneretal:13}. Thus our
method can mask out the contribution arising from the highly-complex
nonlinear physics in a power spectrum measurement, and then allows for
the use of the cleaned lensing power spectrum to do cosmology
\citep[see also][for the similar-idea based
method]{Baldaufetal:10,Mandelbaumetal:12}. 
The method can almost recover the Gaussian information, and we therefore
need not to further measure the higher-order functions of the weak
lensing field to extract the full information of weak lensing.

In this paper, we consider a method of subtracting a theory (or the
data-calibrated) template of the 1-halo term from the measured power
spectrum.  An alternative approach would be to subtract the 1-halo
term contribution cluster by cluster in the two-dimensional shear map. For each halo
region, one can assume an expected shear field around the halo, subtract
the contribution from the measured shear field, and then measure the
power spectrum of the modified shear field. This method may have a
practical advantage in that it can properly take into account variations
in the expected shear field for each halo region. However, our method
in this paper almost recovers the Gaussian information content at
angular scales of interest, and therefore the 1D and 2D based methods
would be almost equivalent -- in other words, there is no significant 
contributions arising from 
the higher-order moments of the shear field around each halo.

However, the results we have shown in this paper are based on several
simplified assumptions. First, we assumed that we can select all massive
halos with masses greater than a sharp mass threshold in the survey
region. In reality, halo mass needs to be inferred from observables,
which therefore involves an unavoidable uncertainty in relating the
observables to halo masses -- scatters and bias in the halo-mass proxy
relation.  An imperfect knowledge of the halo mass proxy causes an
uncertainty in the use of massive halos for cosmology. 
The stacked lensing of sampled halos divided in halo observable bins can
also be used to calibrate the cluster-mass proxy relation, as studied in
\citet{OguriTakada:11}.  In addition, we have ignored possible
systematic errors inherent in weak lensing measurements such as
photometric redshift errors and imperfect shape measurement
\citep[][]{Hutereretal:06,Nishizawaetal:10}.   Hence we
need to further carefully study how the Gaussianization method in this
paper can be applied in the presence of the systematic errors.  In this
paper, we ignored the super-sample variance in the weakly nonlinear
regime, which can be derived based on the perturbation theory
\citep{TakadaHu:13}. The perturbation theory version of the super-sample
variance is not significant compared to other non-Gaussian errors at
scales of interest, as studied in \citet{TakadaJain:09}, but this effect
also needs to be taken into account for an actual application.

The formulation developed in this paper would offer various
applications. Our method is based on the fact that all large-scale
structure probes, drawn from the same survey volume, arise from the same
underlying matter distribution and therefore are correlated with each
other through the super-sample variance effect.  Ideally, we want to
develop a theory to describe the joint likelihood function of all the
observables in order to extract or reconstruct the full information of
the underlying matter field or equivalently the information of the
initial Gaussian field. Since the super-survey modes are not a direct
observable, we need to properly taken into account the super-sample
variance contribution.  Our results suggest that the observed number
counts of massive halos in a given survey volume can be used to
``self-calibrate'' the super-sample variance effect on the power
spectrum measurement or other large-scale structure probes
in the nonlinear regime.  
Our method can be easily extended to the higher-order
functions of matter or weak lensing field and also to weak lensing
tomography. These are our future work and will be presented elsewhere.

\bigskip
 
\section*{Acknowledgments}
We thank Wayne Hu, Bhuvnesh Jain, Issha Kayo, Elisabeth Krause, Tsz Yan
Lam, Roland de Putter, and Emmanuel Schaan for useful discussion and
valuable comments.  We also thank Masanori Sato for providing us with
the ray-tracing simulation data and the halo catalogs used in this work.
MT greatly thanks Department of Astrophysical Sciences, Princeton
University for its warm hospitality during his visit, where this work
was initiated. 
MT also thanks the Aspen Center for Physics and the NSF Grant \#1066293 
and Institut f\"ur Theoretische Physik,
Universit\"at Heidelberg, for their warm hospitality during his visit,
where this work was partly done.  DNS acknowledges support from  the NASA AST theory program
and the US Euclid Science Team.  This work is in part
supported in part by Grant-in-Aid for Scientific Research from the JSPS
Promotion of Science (No. 23340061), 
by Grant-in-Aid for Scientific Research on
Priority Areas No. 467 ``Probing the Dark Energy through an Extremely
Wide \& Deep Survey with Subaru Telescope'', by World Premier
International Research Center Initiative (WPI Initiative), MEXT, Japan,
by the FIRST program ``Subaru Measurements of Images and Redshifts
(SuMIRe)'', CSTP, Japan, and by the exchange program between JSPS and
DFG.

\bibliographystyle{mn2e} \bibliography{mn-jour,refs}

\appendix

\section{Derivation of the power spectrum covariance}
\label{app:pscov}

Here, by using the joint probability distributions of the halo number
counts in a finite-volume survey (Eq.~\ref{eq:p_app}) as well as the
halo model approach, we derive the covariance matrix of the power
spectrum. To do this, we consider the 1- and 2-halo term power spectra
separately, and ignore the cross-correlation for simplicity.

First, let us consider an estimator of the 1-halo term of the power
spectrum: 
\begin{equation}
\hat{P}^{1h}(k)=\frac{1}{V_s}\sum_i \hat N_i \hat{p}^{1h}_i(k).
\end{equation}
Using the probability distribution for the halo number counts, $\left\{
\hat N_1, \hat N_2, \dots, \hat N_i, \dots
\right\}$, the ensemble average of the estimator can be computed as 
\begin{eqnarray}
\ave{\hat{P}^{1h}(k)}&=&\frac{1}{V_s}\sum_i \ave{\hat N_i}
 \ave{\hat{p}^{1h}_i(k)}\nonumber\\
&=& \frac{1}{V_s} \sum_i 
\ave{{\cal L}(\hat N_1, \hat N_2, \dots, \hat N_i,
 \dots)\hat N_i}\ave{ \hat p^{1h}_i(k)}\nonumber\\
&=&  
\frac{1}{V_s}\sum_i 
\sum_{\hat N_1=0}^{\infty}
\sum_{\hat N_2=0}^{\infty}
\cdots
\sum_{\hat N_i=0}^{\infty}
\cdots 
\left[
\prod_{i=1}^k \frac{\bar N_i^{\hat N_i}}{\hat N_i!}\exp(-\bar N_i) 
\right]
\left[
1+\frac{1}{2}\left\{
\left(\sum_jb_j(\hat N_j-\bar{N}_j)\right)^2-\sum_jb_j^2\hat N_j
\right\}
\sigma_m^2
\right] \hat N_i p^{1h}_i (k)\nonumber\\
&=&  
\frac{1}{V_s}\sum_i p^{1h}_i(k)
\sum_{\hat N_i=0}^{\infty}
\frac{\bar{N}_i^{\hat N_i}e^{-\bar{N}_i}}{\hat{N}_i}
\left[
1+\frac{b_i^2}{2}\left\{
(\hat{N}_i-\bar{N}_i)^2-\hat{N}_i
\right\}
\right]\hat{N}_i
\nonumber\\
&=& \frac{1}{V_s}\sum_i \bar{N}_i p^{1h}_i(k)\nonumber\\
&=& \int\!dM\frac{\rmd n}{\rmd M} p^{1h}_M(k).
\end{eqnarray}
Thus the ensemble average of the estimator recovers the 1-halo term
expression in Eq.~(\ref{eq:ps_halomodel}).  Here, in the first line on
the right hand side, we have assumed that the number counts and the halo
profile are independent, and in from the second to fourth lines, we have
used the formula (Eq.~\ref{eq:p_app_form1}).

The covariance matrix of the 1-halo term power spectrum is
defined, as given in \citep{TakadaBridle:07,Kayoetal:13}, as 
\begin{equation}
{\rm Cov}\left[\hat{P}^{1h}(k),\hat{P}^{1h}(k')\right]
\equiv \ave{\hat{P}^{1h}(k)\hat{P}^{1h}(k')}-P^{1h}(k)P^{1h}(k').
\end{equation}
Again, by using the joint probability distribution for the halo number
counts (Eq.~\ref{eq:p_app}), the first term of the above equation can be
computed as 
\begin{eqnarray}
 \ave{\hat{P}^{1h}(k)\hat{P}^{1h}(k')}
&=& \frac{1}{V_s^2}\ave{\sum_i\sum_j \hat N_i\hat N_j
 \hat{p}^{1h}_i(k)\hat{p}^{1h}_j(k')}\nonumber\\
&=& 
\frac{1}{V_s^2}\ave{{\sum_{i}\hat N_i^2}\hat{p}^{1h}_i(k) \hat{p}^{1h}_i(k')}
+\frac{1}{V_s^2}\ave{{\sum_{i,j (i\ne j)}}\hat N_i\hat N_j
\hat{p}^{1h}_i(k) \hat{p}^{1h}_j(k')} 
\nonumber\\
&=& 
\frac{1}{V_s^2}\sum_i 
\sum_{\hat N_i=0}^{\infty}~{\cal L}(\hat N_i)\hat N_i^2 \ave{\hat{p}^{1h}_i(k)\hat{p}^{1h}_i(k')}
+
\frac{1}{V_s^2}\sum_{i,j (i\ne j) }
\sum_{\hat N_i=0}^\infty\sum_{\hat N_{j}=0}^{\infty}~{\cal L}(\hat N_i,\hat N_j)\hat N_i\hat N_j \ave{\hat{p}^{1h}_i(k)\hat{p}^{1h}_j(k')}\nonumber\\
&=&
\frac{1}{V_s^2}\sum_i\left[
\bar{N}_i+\bar{N}_i^2+\bar{N}_i^2b_i^2\sigma_m^2
\right]\left[p^{1h}_i(k)p^{1h}_i(k')+\frac{2}{N_{\rm
mode}(k)}p^{1h}_i(k)^2\delta^K_{kk'}\right] \nonumber\\
&&+\frac{1}{V_s^2}\sum_{i,j (i\ne j)}
\left[\bar{N}_i\bar{N}_j + b_ib_j\bar{N}_i\bar{N}_j\sigma_m^2
\right]\left[p^{1h}_i(k)p^{1h}_i(k')+\frac{2}{N_{\rm
mode}(k)}p^{1h}_i(k)p^{1h}_j(k)\delta^K_{kk'}\right]\nonumber\\
&&\hspace{-4em}=
P^{1h}(k)P^{1h}(k')+
\frac{1}{V_s}\int\!dM\frac{\rmd n}{\rmd M}p^{1h}_M(k)p^{1h}_M(k')
+
\frac{1}{V_s}\frac{2}{N_{\rm mode}(k)}\delta^K_{kk'}
\int\!dM\frac{\rmd n}{\rmd M}p^{1h}_M(k)^2
+\frac{2}{ N_{\rm mode}(k)}P^{1h}(k)^2\delta^K_{kk'}
\nonumber\\
&&
+\left(1+\frac{2}{\hat N_{\rm mode}(k)}\delta^K_{kk'}\right)
\left[
\int\!dM\frac{\rmd n}{\rmd M}b(M)p^{1h}_M(k)
\right]
\left[
\int\!dM'\frac{\rmd n}{\rmd M'}b(M')
p^{1h}_{M'}(k')\right]\sigma_m^2,
\end{eqnarray}
where we have used the summation rules (Eq.~\ref{eq:p_app_form1}) and
also used the following forms to convert the summation and integration
forms
\begin{eqnarray}
&&P^{1h}(k)\equiv \int\!dM~\frac{\rmd n}{\rmd M}p^{1h}_M(k)\simeq 
\frac{1}{V_s}\sum_{i}\bar{N}_i p^{1h}_i(k),\\
&&T^{1h}(k,k')\equiv \int\!dM~\frac{\rmd n}{\rmd M}p^{1h}_M(k)p^{1h}_M(k')
\simeq \frac{1}{V_s}\sum_{i}\bar{N}_i p^{1h}_i(k)p^{1h}_i(k'),\\
&&\int\!dM~\frac{\rmd n}{\rmd M}b(M)p^{1h}_M(k)
\int\!dM'\frac{\rmd n}{\rmd M'}b(M')p^{1h}_{M'}(k)\simeq \frac{1}{V_s^2}
\sum_{i,j}\bar{N}_i\bar{N}_{j}b_ib_j
p^{1h}_i(k)p^{1h}_j(k').
\end{eqnarray}
Therefore, the covariance matrix of the 1-halo term power spectrum is
given as
\begin{eqnarray}
{\rm Cov}[\hat P^{1h}(k),\hat P^{1h}(k')]&\simeq & 
\frac{2}{ N_{\rm mode}(k)}P^{1h}(k)^2\delta^K_{kk'}
+\frac{1}{V_s}T^{1h}(k,k')
+\left[
\int\!dM\frac{\rmd n}{\rmd M}b(M)p^{1h}_M(k)
\right]
\left[
\int\!dM'\frac{\rmd n}{\rmd M'}b(M')
p^{1h}_{M'}(k')\right]\sigma_m^2.\nonumber\\
\label{eq:cov_1h}
\end{eqnarray}
The third term is the halo sample variance term found in
\cite{Satoetal:09} \citep[also see][]{Kayoetal:13}. 

Similarly, we can work on an estimator of the 2-halo term power
spectrum. The ensemble average of the 2-halo term can be computed as 
\begin{eqnarray}
\ave{\hat{P}^{2h}(k)}&=&\frac{1}{V_s^2}
\sum_{i,j(i\ne j)}\ave{\hat N_i\hat N_j}\ave{\hat{p}^{2h}_{ij}(k)}\nonumber\\
&=&\frac{1}{V_s^2}\sum_{i,j (i\ne j)}
\left[\bar{N}_i\bar{N}_j + b_ib_j\bar{N}_i\bar{N}_j\sigma_m^2
\right]p_{ij}^{2h}(k)\nonumber\\
&=&\int\!dM\frac{\rmd n}{\rmd M} \int\!dM'\frac{\rmd n}{\rmd M'} p_{MM'}^{2h}(k)
+\sigma_m^2\int\!dM\frac{\rmd n}{\rmd M}b(M) 
\int\!dM'\frac{\rmd n}{\rmd M'}b(M') p_{MM'}^{2h}(k).
\end{eqnarray}
The covariance matrix is computed, up to the order of $\sigma_m^2$, as
\begin{eqnarray}
{\rm Cov}[\hat{P}^{2h}(k),\hat{P}^{2h}(k')]&=&
\frac{1}{V_s^4}\ave{\sum_{i,j (i\ne j)}\sum_{i',j' (i'\ne j')}
\hat N_i\hat N_j \hat N_{i'}\hat N_{j'}
\hat{p}^{2h}_{ij}(k)\hat{p}^{2h}_{i'j'}(k')}
-P^{2h}(k)P^{2h}(k')
\nonumber\\
&&\hspace{-6em}=\frac{1}{V_s^4}\sum_{i,j}\sum_{i',j'}\ave{
\hat N_i \hat N_j \hat N_{i'} \hat N_{j'}}\ave{\hat
p^{2h}_{ij}(k)\hat{p}^{2h}_{i'j'}(k')}-P^{2h}(k)P^{2h}(k')\nonumber\\
&&\hspace{-6em}=
\frac{1}{V_s^4}\sum_{i,j}\sum_{i',j'}
\bar{N}_i\bar{N}_j\bar{N}_{i'}\bar{N}_{j'}
\left[1+\sigma_m^2\left\{
b_ib_j+b_ib_{i'}+b_ib_{j'}+b_jb_{i'}+b_{j}b_{j'}+b_{i'}b_{j'}
\right\}
\right]\nonumber\\
&&\hspace{-3em}\times
\left[p^{2h}_{ij}(k)p^{2h}_{i'j'}(k')
+\frac{1}{N_{\rm mode}(k)}\delta^K_{kk'}\left(
V_st_{ii'jj'}(k,k')+
p_{ii'}^{2h}(k)p^{2h}_{jj'}(k)+p^{2h}_{ij'}(k)p^{2h}_{ji'}(k)
\right)
\right]-P^{2h}(k)P^{2h}(k')\nonumber\\
&&\hspace{-6em}=
\frac{1}{V^3_s}\sum_{i,j,i',j'}
\bar{u}_i\bar{u}_j\bar{u}_{i'}\bar{u}_{j'}
t_{iji'j'}(k,k')
+\frac{1}{V_s^4}\sum_{i,j}\sum_{i',j'}
\bar{N}_i\bar{N}_j\bar{N}_{i'}\bar{N}_{j'}\sigma_m^2\left[
b_{i}b_{i'}+b_ib_{j'}+b_{j}b_{i'}+b_{j}b_{j'}+b_{i'}b_{j'}
\right]p_{ij}^{2h}(k)p_{i'j'}^{2h}\nonumber\\
&&\hspace{-4em}
+\frac{1}{V_s^4}\sum_{i,j}\sum_{i',j'}
\bar{N}_i\bar{N}_j\bar{N}_{i'}\bar{N}_{j'}
\frac{1}{N_{\rm mode}(k)}\delta^K_{kk'}\left[
p_{ii'}^{2h}(k)p_{jj'}^{2h}(k)
+p_{ij'}^{2h}(k)p_{ji'}^{2h}(k)
\right]\nonumber\\
&&\hspace{-6em}=\frac{1}{V_s}\bar{T}^{4h}(k,k')
+\frac{2}{N_{\rm mode}(k)}\delta^{K}_{kk'}P^{2h}(k)^2
\nonumber\\
&&\hspace{-2em}
+
4\sigma_m^2\int\!\!\left[\prod_{i=1}^4dM_i \frac{\rmd n}{\rmd M_i}\right]
\left[
b_1b_2+b_1b_3+b_1b_4+b_2b_3+b_3b_4
\right]
p^{2h}_{M_1M_2}(k)p^{2h}_{M_3M_4}(k'),
\label{eq:cov_2h}
\end{eqnarray}
where $T^{4h}$ is the 4-halo term of the trispectrum, and we have
introduced notational convention such as $b_1=b(M_1)$. 

Hence, summarizing Eqs.~(\ref{eq:cov_1h}) and (\ref{eq:cov_2h}), 
the covariance matrix for the power spectrum is given as 
\begin{eqnarray}
{\rm Cov}[\hat{P}(k),\hat{P}(k')]&=&
\frac{2}{N_{\rm
mode}(k)}P(k)^2\delta^{K}_{kk'}+\frac{1}{V_s}\bar{T}(k,k')\nonumber\\
&&+\sigma_m^2
 \int\!\!dM\frac{\rmd n}{\rmd M}b(M)p^{1h}_M(k)\int\!\!dM\frac{\rmd
 n}{\rmd M}b(M)p^{1h}_{M'}(k')\nonumber\\
&&+4\sigma_m^2\int\!\!\left[\prod_{i=1}^4dM_i \frac{\rmd n}{\rmd M_i}\right]
\left[
b_1b_2+b_1b_3+b_1b_4+b_2b_3+b_3b_4
\right]
p^{2h}_{M_1M_2}(k)p^{2h}_{M_3M_4}(k').
\end{eqnarray}
The two terms on the first line of the r.h.s. are the standard
covariance terms, and the other 2 terms are due to the halo sample
variance. The fourth term (the last term) on the r.h.s. is much smaller
than the third term  at scales of interest, and therefore we ignore the
fourth term in main text. 

\section{Cross-correlation between the halo counts and the power
 spectrum}
\label{app:cross}

The cross-correlation between the halo number counts and the 1-halo term
power spectrum can be computed as
\begin{eqnarray}
{\rm Cov}[\hat N_{i'},\hat{P}^{1h}(k)]&\equiv &
 \ave{\hat N_{i'}\hat{P}^{1h}(k)}- \ave{\hat N_{i'}}P^{1h}(k)
\nonumber\\
&=& \frac{1}{V_s}\sum_i
 \ave{\hat N_{i'}\hat N_i}\ave{\hat{p}^{1h}_i(k)}-\ave{\hat N_{i'}}P^{1h}(k)
\nonumber\\
&=&\frac{1}{V_s}\left[\ave{\hat N_{i'}^2}p_{i'}^{1h}(k)
+\sum_{i (i\ne i')}\hat N_{i'}\hat N_ip^{1h}_i(k)\right]-\bar{N}_{i'}P^{1h}(k)\nonumber\\
&=&\frac{1}{V_s}\left[
\sum_{\hat N_{i'}=0}^{\infty}{\cal L}(\hat N_{i'})\hat N_{i'}^2 p_{i'}^{1h}(k)
+\sum_{i (i\ne i')}\sum_{\hat N_i=0}^\infty \sum_{\hat N_{i'}=0}^\infty{\cal
L}(\hat N_i,\hat N_{i'}) \hat N_{i'}\hat N_i p^{1h}_i(k)
\right]-\bar{N}_{i'}P^{1h}(k)\nonumber\\
&=& \frac{1}{V_s}\left[
(\bar{N}_{i'}+\bar{u}_{i'}^2+
{b}_{i'}^2\bar{N}_{i'}^2\sigma_m^2)p_{i'}^{1h}(k)
+\sum_{i (i\ne i')}
\bar{N}_{i'}\bar{N}_{i}(1+b_{i'}b_{i}\sigma_m^2)p^{1h}_i(k)
\right]-\bar{N}_{i'}P^{1h}(k)\nonumber\\
&=& \frac{1}{V_s}\bar{N}_{i'}p_{i'}^{1h}(k)+
\frac{1}{V_s}b_{i'}\bar{N}_{i'} \sigma_m^2\sum_{i}b_i \bar{N}_i
p_{i}^{1h}(k )\nonumber\\
&=& \frac{1}{V_s}\bar{N}_{i'}p_{i'}^{1h}(k)+
b_{i'}\bar{N}_{i'} \sigma_m^2\int\!\!dM\frac{\rmd n}{\rmd M} b(M)p_M^{1h}(k).
\end{eqnarray}
Similarly, the cross-covariance between the halo number counts and the
2-halo term power spectrum is 
\begin{eqnarray}
{\rm Cov}[\hat N_{i'},\hat{P}^{2h}(k)]&\equiv &
 \ave{\hat N_{i'}\hat{P}^{2h}(k)}- \ave{\hat N_{i'}}P^{2h}(k)
\nonumber\\
&=& \frac{1}{V_s^2}\sum_{j,k}\hat{N}_i\hat{N}_j\hat{N}_k
\ave{ \hat{p}_{jk}^{2h}(k)} -\bar{N}_iP^{2h}(k)\nonumber\\
&=&\frac{1}{V_s^2}\sum_{j,k}
\bar{N}_i\bar{N}_j\bar{N}_k
\left[1+\sigma_s^2\left(b_ib_j+b_ib_k+b_jb_k\right)\right]p^{2h}_{jk}(k)-\bar{N}_iP^{2h}(k)\nonumber\\
&=&\bar{N}_i\sigma_m^2
\int\!dMdM'\frac{\rmd n}{\rmd M}\frac{\rmd n}{\rmd M'}\left[
b(M_i)b(M)+b(M_i)b(M')+b(M)b(M')
\right]p^{2h}_{MM'}(k).
\end{eqnarray}

Therefore, the cross-covariance between the halo number counts and the
power spectrum is given as
\begin{eqnarray}
{\rm Cov}[\hat{N}_i,\hat{P}(k)]&=& \frac{1}{V_s}\bar{N}_{i'}p_{i'}^{1h}(k)+
b_{i'}\bar{N}_{i'} \sigma_m^2\int\!\!dM\frac{\rmd n}{\rmd M} b(M)p_M^{1h}(k)
\nonumber\\
&&+\bar{N}_i\sigma_m^2
\int\!dMdM'\frac{\rmd n}{\rmd M}\frac{\rmd n}{\rmd M'}\left[
b(M_i)b(M)+b(M_i)b(M')+b(M)b(M')
\right]p^{2h}_{MM'}(k).
\end{eqnarray}

\end{document}